\documentclass{aa}

\usepackage[varg]{txfonts}
\usepackage{natbib}
\usepackage{graphicx}
\usepackage{xspace}
\usepackage{bm}
\usepackage[flushleft]{threeparttable}

\defcitealias{Soler18}{Soler18}
\newcommand{\g}{$\gamma$\xspace}
\newcommand{\halpha}{H$_{\alpha}$\xspace}
\newcommand{\wco}{$W_{\rm{CO}}$\xspace}

\newcommand{\hi}{\ion{H}{i}\xspace}
\newcommand{\hii}{\ion{H}{ii}\xspace}
\newcommand{\nh}{$N_{\rm{H}}$\xspace}
\newcommand{\nhi}{$N_{\ion{H}{i}}$\xspace}

\newcommand{\nhtot}{$N_{\rm{Htot}}$\xspace}

\newcommand{\wcounit}{K\,km\,s$^{-1}$\xspace}

\newcommand{\percc}{cm$^{-3}$\xspace}

\newcommand{\kmpers}{km\,s$^{-1}$\xspace}
\newcommand{\southloop}{south loop\xspace}
\newcommand{\northrim}{north rim\xspace}
\newcommand{\eastshell}{east shell\xspace}
\newcommand{\westrim}{west rim\xspace}
\newcommand{\bsky}{$B_{\rm sky}$\xspace}
\newcommand{\blos}{$B_{\rm los}$\xspace}
\newcommand{\btot}{$B_{\rm tot}$\xspace}

\def\Planck{\textit{Planck}\xspace}
\def\ROSAT{\textit{ROSAT}\xspace}

\begin{document}

\title{Gas shells and magnetic fields in the Orion-Eridanus superbubble}

\author{T. Joubaud\inst{1}
\and I. A. Grenier\inst{1}
\and J. Ballet\inst{1}
\and J. D. Soler\inst{2}}

\institute{
\inst{1}~Laboratoire AIM, CEA-IRFU/CNRS/Universit\'e Paris Diderot, D\'epartement d’Astrophysique, CEA Saclay, F-91191 Gif sur Yvette, France \\
\inst{2}~Max-Planck-Institute for Astronomy, K\"onigstuhl l17, 69117 Heidelberg, Germany \\
\email{theo.joubaud@cea.fr} \\
\email{isabelle.grenier@cea.fr} \\
\email{soler@mpia.de}
}

\date{Received 05/07/2019 /
Accepted 22/08/2019}

\abstract {}
{The Orion-Eridanus superbubble has been blown by supernovae and supersonic winds of the massive stars in the Orion OB associations. It is the nearest site at which stellar feedback on the interstellar medium that surrounds young massive clusters can be studied. 
The formation history and current structure of the superbubble are still poorly understood, however. It has been pointed out that the  picture of a single expanding object should be replaced by a combination of nested shells that are superimposed along the line of sight. We have investigated the composite structure of the Eridanus side of the superbubble in the light of a new decomposition of the atomic and molecular gas.}
{We used \hi 21 cm and CO (J=1$-$0) emission lines to separate coherent gas shells in space and velocity, and we studied their relation to the warm ionised gas probed in \halpha emission, the hot plasma emitting X-rays, and the magnetic fields traced by dust polarised emission. We also constrained the relative distances to the clouds using dust reddening maps and X-ray absorption. We applied the Davis-Chandrasekhar-Fermi method to the dust polarisation data to estimate the plane-of-sky components of the magnetic field in several clouds and along the outer rim of the superbubble.}
{Our gas decomposition has revealed several shells inside the superbubble that span distances from about 150 pc to 250 pc. One of these shells forms a nearly complete ring filled with hot plasma. Other shells likely correspond to the layers of swept-up gas that is compressed behind the expanding outer shock wave. We used the gas and magnetic field data downstream of the shock to derive the shock expansion velocity, which is close to $\sim$20~\kmpers. Taking the X-ray absorption by the gas into account, we find that the hot plasma inside the superbubble is over-pressured compared to plasma in the Local Bubble. The plasma comprises a mix of hotter and cooler gas along the lines of sight, with temperatures of (3-9) and $(0.3-1.2)\times 10^6$~K, respectively. The magnetic field along the western and southern rims and in the approaching wall of the superbubble appears to be shaped and compressed by the ongoing expansion. We find plane-of-sky magnetic field strengths from 3 to 15~$\mu$G along the rim.}
{}


\maketitle

\section{Introduction}
\begin{figure}
\resizebox{\hsize}{!}{\includegraphics{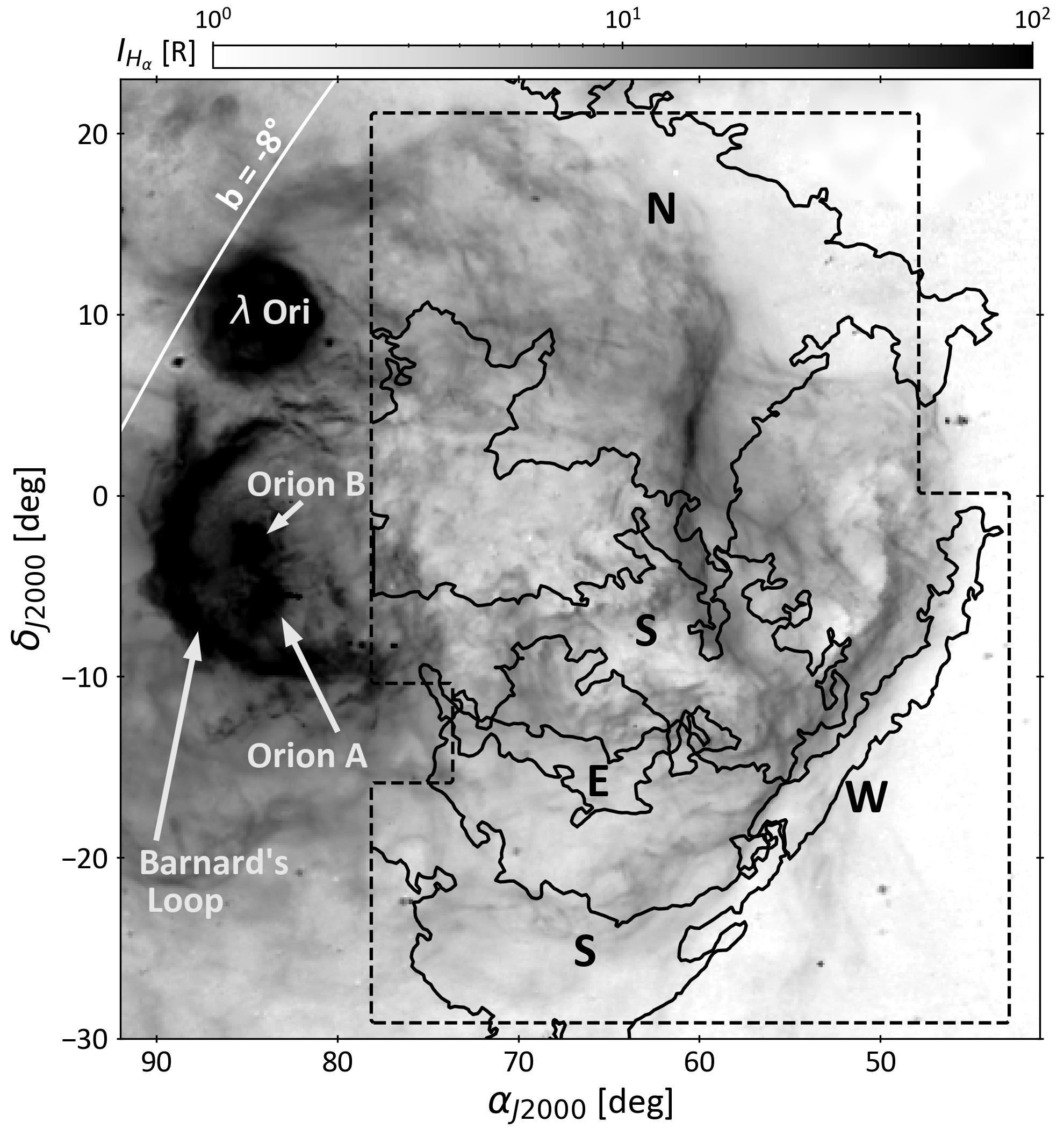}}
\caption{\halpha intensity map of the Orion-Eridanus superbubble based on VTSS, SHASSA, and WHAM data. The dashed line traces the contour of the analysed region. The white labels on the left-hand side show the key \halpha features towards Orion. In the analysed region, the black contours delineate the main \hi shells we identify in Sect. \ref{sec:HICO}: the \northrim (N), \southloop (S), \eastshell (E), and \westrim (W). The line of constant Galactic latitude b=-8\degr\ in the upper left corner indicates the orientation of the Galactic plane.}
\label{fig:Halpha}
\end{figure}

The Orion-Eridanus superbubble is the nearest site of active high-mass star formation. It has been studied in particular to investigate the feedback of these stars on the interstellar medium. Together with supernovae and stellar winds, the intense UV stellar radiation has carved a 200 pc wide cavity that is filled with low-density ionised gas and expands into the interstellar medium. The material that was initially present was swept up and compressed to form a shell of neutral gas, the near wall of which may be as close as 180 pc from the Sun \citep{Bally2008}. The boundary between the ionised gas and the neutral shell can be seen in \halpha emission that traces the recombination of ionised hydrogen. The \halpha emission that is detected in the region is displayed in Fig. \ref{fig:Halpha}. This figure features the \hii region $\lambda$ Orionis, the bright crescent of Barnard's Loop close to the Orion A and B molecular clouds, and the arcs of the Eridanus Loop to the west. The westernmost arc nicely delineates the outer rim of the superbubble, but the origin of the brightest vertical arc, called Arc A by \citet{Johnson78}, is still debated. It may come from a shell inside the superbubble or from the superposition of separate structures along the line of sight. \citet{Pon14} summarised arguments in favour and against the association of Arc A with the superbubble based on several tracers and methods, but did not firmly conclude. They mentioned that the molecular cloud identified as MBM 18 (for Magnani, Blitz and Mundi, near $\alpha_{J2000}=60\fdg6, \delta_{J2000}=1\fdg3$) by \citet{Magnani1985} coincides in direction and tentatively in velocity with Arc A. The detection of dust reddening fronts in the PanSTARRS-1 stellar photometric data place MBM 18 within 200 pc from the Sun \citep{Zucker2019}. No significant reddening is found beyond 500 pc in this direction; this strengthens the association between Arc A and the superbubble.

The H$_\alpha$ emission has also been used to determine the outer boundaries of the superbubble. Whereas it appears rather clearly defined on the western side, it had been thought that the eastern boundary was Barnard's Loop until \citet{Ochsendorf15} pushed the limit farther out to a fainter feature beyond the extent of Figure \ref{fig:Halpha}. The authors argued that the dusty shell of Barnard's Loop was not optically thick enough to absorb all ionising photons from the Orion OB1 association. These photons leak through to a more distant wall that is identified in gas velocity and as a faint \halpha filament closer to the Galactic plane. Barnard's Loop is part of a closed bubble that may be driven by a supernova and expands inside the superbubble. The  other shells (GS206-17+13, Orion Nebula) and the complex history of $\lambda$ Orionis \citep[possibly a supernova remnant cavity that has later been filled by the \hii region around the OB association,][]{Dolan02} led \citet{Ochsendorf15} to change the simple picture of a single expanding object to a more complex combination along the line of sight of evaporating clouds and nested shells filled with X-ray emitting hot gas at different temperatures \citep{Snowden95}.

On the plane of the sky, the elongated bubble extends from beyond Orion up to the Eridanus Loop, but its overall orientation is still uncertain. Absorption line measurements \citep{Welsh05} and dust extinction data from Gaia and the two micron all-sky survey (2MASS) \citep{Lallement18} support a close end towards Eridanus (southwest) and a far end beyond Orion (east), whereas models of a single shock front expanding in an exponentially stratified interstellar medium (ISM) fit the \halpha data better, with the Orion side being closer than the Eridanus side \citep{Pon16}. The models cannot accommodate the composite structure of the superbubble, however, which has likely evolved in space and time from near to far along the blue stream of massive stars identified by \citet{Pellizza05} and \citet{Bouy15} between us and the Orion clouds.

We have performed a multi-wavelength analysis of the gas and cosmic-ray content of the Eridanus side of the superbubble, which we present in a series of three papers. This paper focuses on the composite structure of atomic and molecular gas shells and on their relation to the hot ionised gas, the recombination fronts, and the magnetic field structure. The second paper (Joubaud et al., in prep., hereafter Paper II) explores the cosmic-ray flux that pervades the gas shells. The third paper (Joubaud et al., in prep.) studies the gas mass found at the transition between the atomic and molecular phases and in the CO-bright parts, as well as the evolution of the dust emission opacity (per gas nucleon) across the different gas phases. The third paper also studies the small molecular cloud of MBM 20, also called Lynd dark nebula (LDN) 1642, that might be compressed at the edge of the Local and/or Eridanus bubbles, near $\alpha_{J2000}=68\fdg8, \delta_{J2000}=-14\fdg2$.

Here we investigate the composite structure of the Orion-Eridanus superbubble in the light of a new  separation of the \hi gas in position and velocity. We  find new shells and study their relation to dust reddening fronts, \halpha and X-ray emissions, and the magnetic field orientation derived from dust-polarised emission. We also use the \hi line information to estimate the magnetic field intensity in several parts of the superbubble.

The paper is structured as follows. Data are presented in Sect. \ref{sec:data}. The separation of the \hi and CO clouds and their relations in velocity, distance, and to known entities are discussed in Sect. \ref{sec:HICO}. We study the X-ray emission and derive the emitting plasma properties in Sect \ref{sec:X-ray}. We probe the magnetic field orientation and strength in Sect. \ref{sec:Bfield}, before we conclude in Sect. \ref{sec:concl}. We present X-ray optical depth maps in Appendix \ref{ap:tauX} and the two methods we used to derive the angular dispersion of the magnetic field in Appendix \ref{ap:polar}.

\begin{figure*}
\centering
\includegraphics[width=17cm]{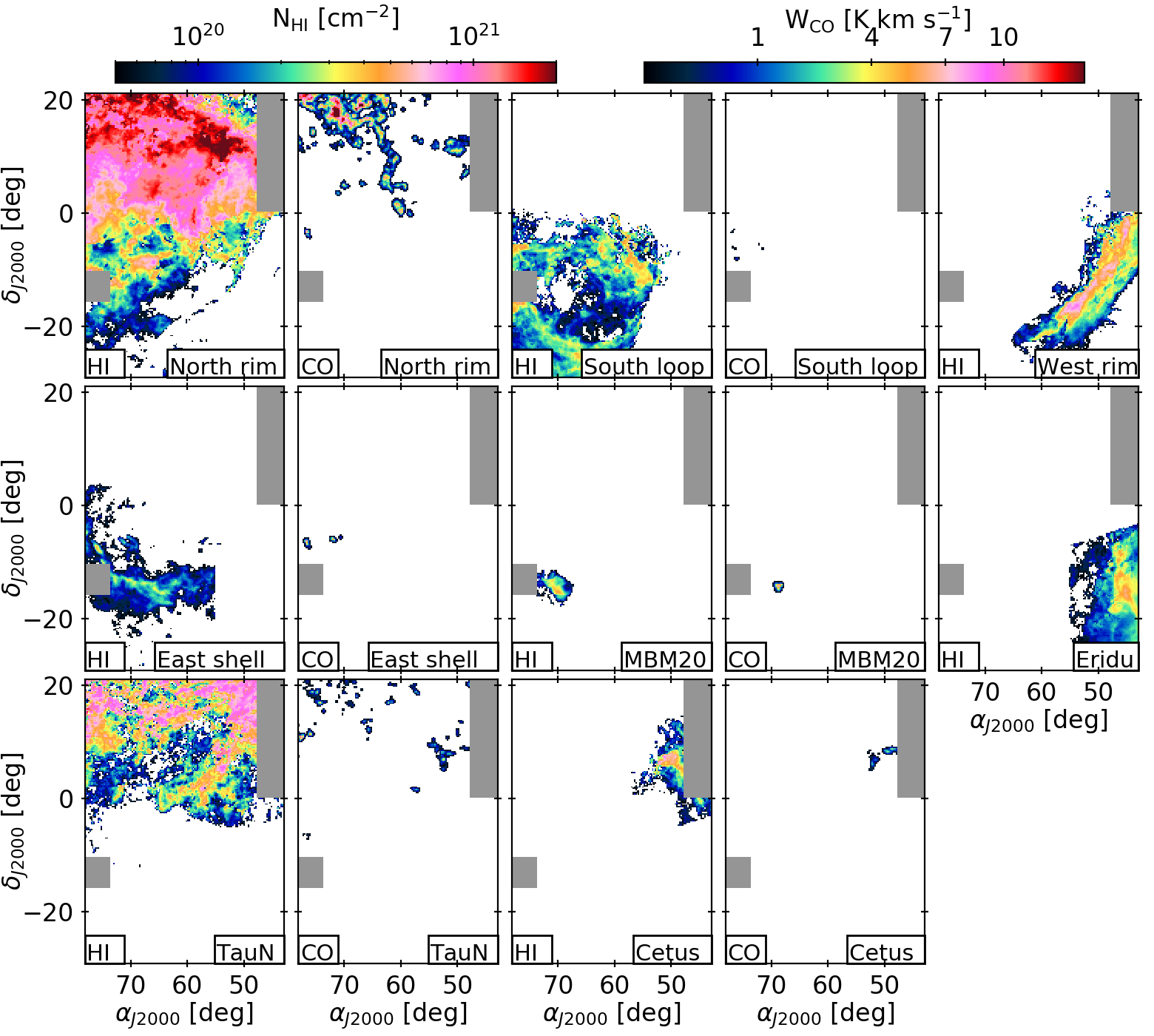}
\caption{Maps of the \northrim, \southloop, \eastshell, \westrim, MBM 20, Eridu cirrus, Cetus, and North Taurus components in \hi column density, \nhi for a spin temperature of 100~K, and in CO line intensity, \wco.}
\label{fig:DustModel}
\end{figure*}

\section{Data}
\label{sec:data}

We have analysed the eastern part of the superbubble towards the Eridanus constellation. It extends from 43\degr\ to 78\degr\ in right ascension and from -29\degr\ to 21\degr\ in declination, as shown in Fig. \ref{fig:Halpha}. We masked two 5\degr\  wide areas on the western and eastern sides of the region to avoid complex gas distributions in the background. All maps are projected onto the same 0.25$^\circ$ spaced Cartesian grid.

        \subsection{Ionised gas}
Warm ionised gas is visible through \halpha emission. It is displayed in Fig. \ref{fig:Halpha} using the data of \citet{Finkbeiner2003}. This is a composite map of the Virginia Tech Spectral line Survey (VTSS), the Southern H-Alpha Sky Survey Atlas (SHASSA), and the Wisconsin H-Alpha Mapper (WHAM). The velocity resolution is 12 km s$^{-1}$ and the spatial resolution is 6\arcmin\ (full width at half maximum, FWHM).

        \subsection{\hi and CO emission line data}
We used the 16'2 resolution HI4PI survey \citep{HI4PI}, with a velocity resolution of 1.49 km/s in the local standard of rest (LSR). We selected velocities between -90 and 50 km s$^{-1}$ to exclude the \hi emission from the high-velocity clouds that lie in the hot Galactic corona far behind the local medium we are interested in \citep{Wakker2008}.

In order to trace the molecular gas, we used the 8\farcm5 resolution $^{12}$CO (J=1$-$0) observations at 115 GHz from the moment-masked CfA CO survey \citep{Dame2001, Dame2004}. We completed this dataset with the CO observations of the MBM 20 cloud obtained with the Swedish-ESO Submillimetre Telescope (SEST) that were kindly provided by D. Russeil \citep{Russeil2003}.

        \subsection{X-ray data}
In order to study the hot-gas content of the superbubble, we used data from the Roentgen satellite (ROSAT) X-ray all-sky survey \citep{Snowden94}. The spatial resolution is about 30\arcsec\ on-axis. We combined the original energy bands into three bands: the 0.25 keV band, spanning from 0.11 to 0.284 keV at 10\% of peak response  (R1+R2), the 0.75 keV band from 0.44 to 1.2 keV (R4+R5), and the 1.5 keV band from 0.7 to 2 keV (R6+R7). The exposure over the analysed region varies between 0 and 600~s, therefore we masked out the underexposed zones with exposure times below 80~s. They appear as white stripes in our intensity maps. Even though we used count rates that take striped variations in the exposure map into account, we caution that systematic biases remain in the survey calibration. They appear as striped enhancements or diminutions in count rates that cross the analysis region roughly parallel to the underexposed stripes. We re-sampled the maps onto our 0\fdg25 grid and have compared the two low-energy count rate maps with those of \citet{Snowden94}.

        \subsection{Dust data}
We used the 3D dust reddening maps of \citet{Green18}. These maps span three quarters of the sky ($\delta\gtrsim$-30\degr) and are based on the stellar photometry of 800 million stars from Pan-STARRS 1 and 2MASS. The authors divided the sky into pixels containing a few hundred stars each. This results in a map with pixel sizes that vary from 3\farcm4 to 55\arcmin.

The polarisation data, that is, the total intensity and the $Q$ and $U$ Stokes parameters, come from the \Planck 353\,GHz 2018 polarisation data with an original angular resolution of 5\arcmin\ \citep{PlanckXII18}. To enhance the signal-to-noise ratio, we smoothed the maps and their covariances to 40\arcmin\ resolution (FWHM) and downgraded the healpix maps to $N_{side}=512$.

\section{\hi and CO cloud separation}
\label{sec:HICO}

        \subsection{Velocity decomposition}

Following the method developed by \citet{PlanckXXVIII15}, we decomposed the \hi and CO velocity spectra into individual lines and used this information to identify and separate eight nearby cloud complexes that are coherent in position and in velocity. We thus fitted each \hi or CO spectrum as a sum of lines with pseudo-Voigt profiles that combine a Gaussian and a Lorentzian curve that share the same mean and standard deviation in velocity. The Lorentzian part of the profile can adapt, if necessary, to extended line wings that are broadened by velocity gradients within the beam. The prior detection of line peaks and shoulders in each spectrum provided a limit on the number of lines to be fitted as well as initial guesses for their central velocities. We improved the original method by taking the line information of the neighbouring spectra into account for each direction in order to better trace merging or fading lines. The fits yield small residuals between the modelled and observed spectra, but in order to preserve the total intensity that is recorded in each spectrum, we distributed the residuals among the fitted lines according to their relative strength in each channel. 

\begin{figure}
\resizebox{\hsize}{!}{\includegraphics{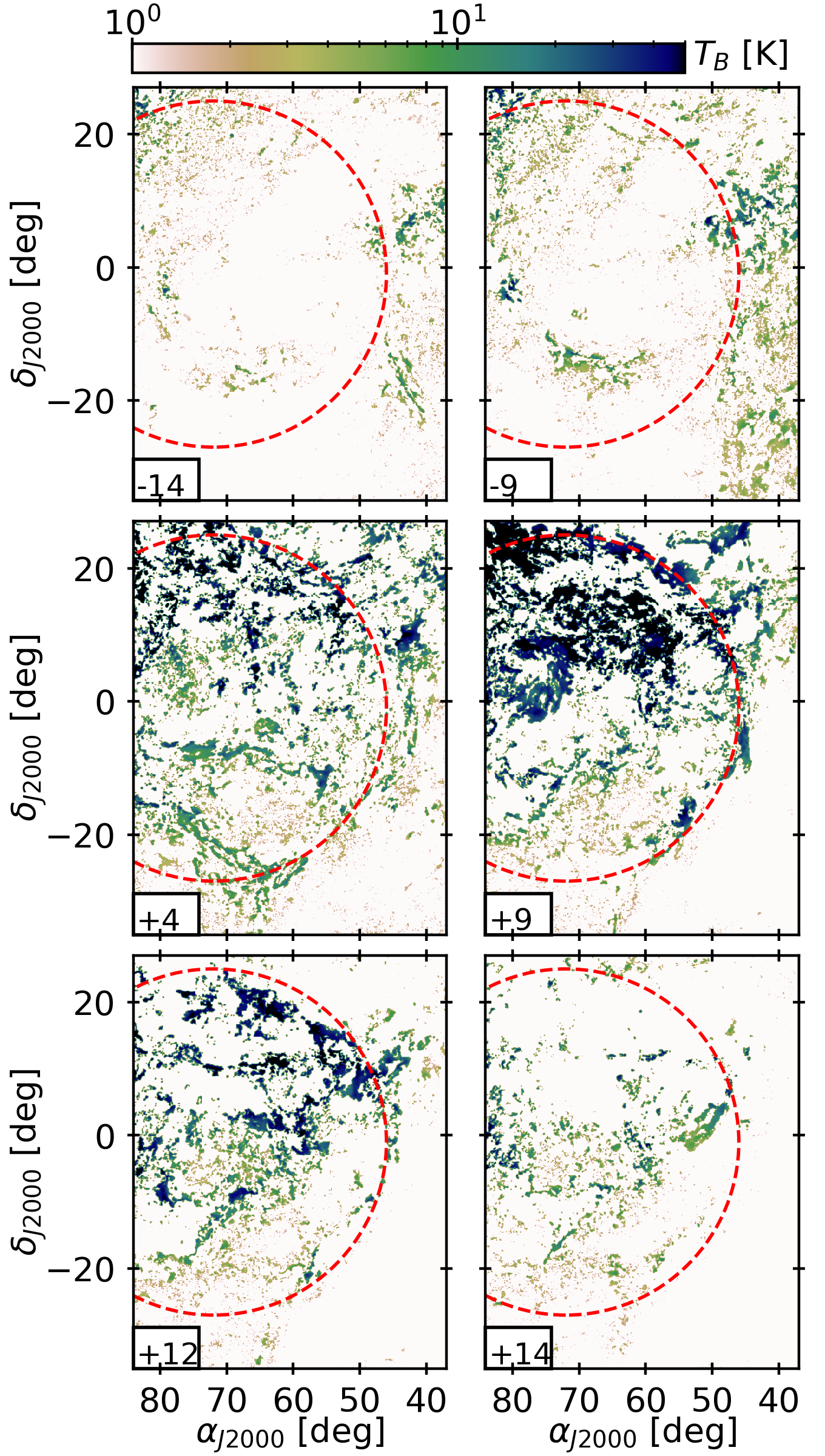}}
\caption{Spatial distribution of the peak temperatures of the fitted \hi lines (grey scale) for different slices in central velocity (indicated in \kmpers in the lower left corner of each plot). The red circle highlights the outer shell that is visible at +9 \kmpers.}
\label{fig:HI9kms}
\end{figure}

\begin{figure*}[ht!]
\centering
\includegraphics[width=17cm]{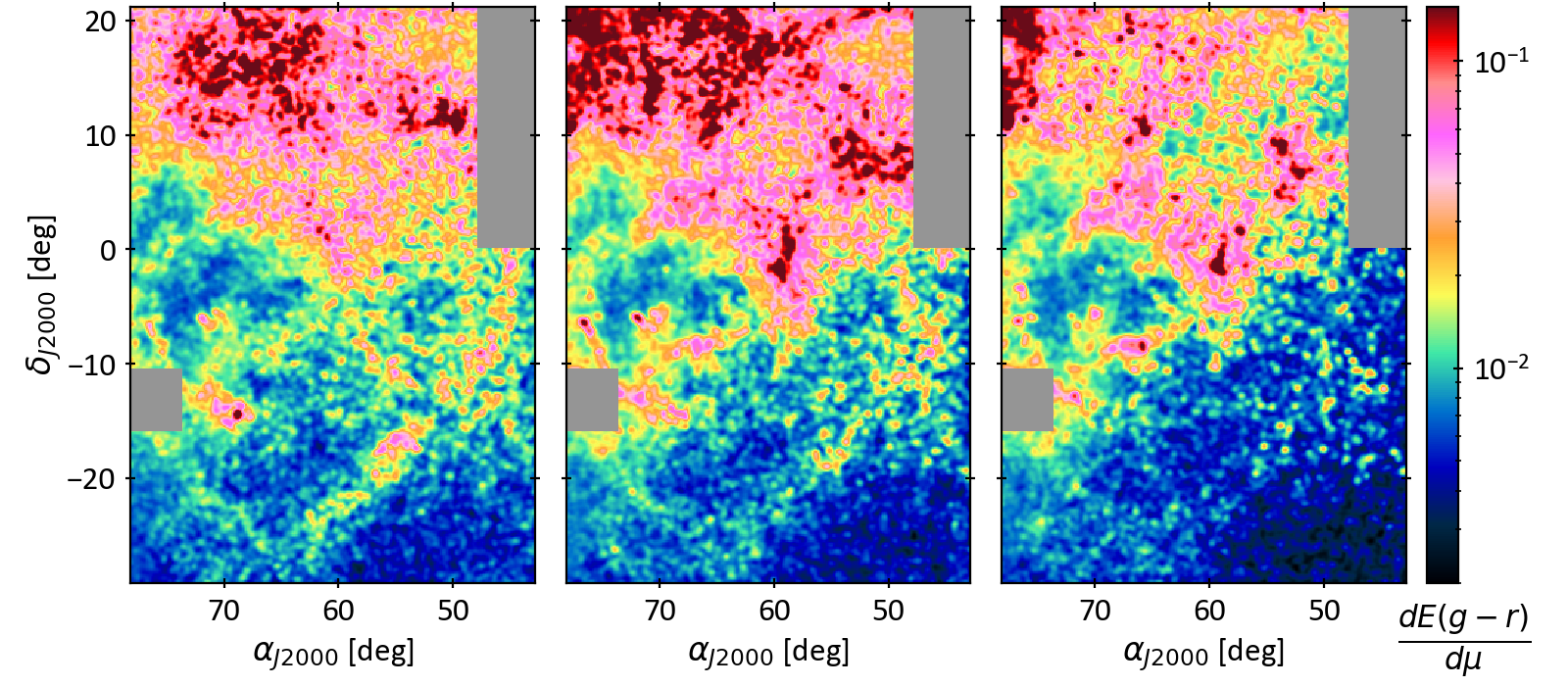}
\caption{Maps of the dust reddening per unit distance modulus, $\frac{dE(g-r)}{d\mu}$, showing the morphology of dust fronts at distances of 200 pc (\textit{left}), 320 pc (\textit{middle}), and 400 pc (\textit{right}). The maps are smoothed by a 0\fdg4 Gaussian kernel for display.}
\label{fig:DistMap}
\end{figure*}

    \subsection{\hi and CO components}
The 3D (right ascension, declination, and central velocity) distribution of the peak temperatures of the fitted lines showed that the data can be partitioned into several entities that are depicted in \nhi column densities and in \wco intensities in Fig. \ref{fig:DustModel}. Some were detected in both \hi and CO emission lines, such as the \northrim, \southloop, \eastshell, Cetus, North Taurus, and MBM 20. The others were only seen in the \hi data, such as the \westrim and Eridu cirrus. To construct these maps, we defined 3D boundaries in right ascension, declination, and velocity for each component. The velocity range for each cloud is presented in Table \ref{tab:Disttable}. We selected the fitted lines with central velocities falling within the appropriate velocity interval, depending on the ($\alpha, \delta$) direction, and we integrated their individual profiles in velocity. The resulting maps were then re-sampled into the 0\fdg25 spaced Cartesian grid of the analysed region.

In order to investigate the effect of the unknown \hi optical depth, we derived all the \nhi maps for a set of uniform spin temperatures (100, 125, 200, 300, 400, 500, 600, 700, and 800~K) and for the optically thin case. \citet{Nguyen18} have recently shown that a simple isothermal correction of the emission spectra with a uniform spin temperature reproduces the more precise \hi column densities that are inferred from the combination of emission and absorption spectra quite well. Their analysis fully covers the range of column densities in the Eridanus region. All the figures and results presented here were derived for a spin temperature of 100~K. This choice comes from the best-fit models of the \g-ray data and of the dust optical depth at 353\,GHz; they are detailed in Paper II. This choice has no effect on cloud separation. Compared with the optically thin case, the correction increases the highest column densities ($> 1.5 \times 10^{21}$~cm$^{-2}$) of the thickest cloud of the \northrim\  by 40\%. The other clouds are more optically thin.  

As we discuss in this paper, some structures are likely associated with the superbubble (\northrim, \southloop, \eastshell, and \westrim), while the others are foreground (MBM 20), background (Cetus, North Taurus) or side (Eridu) clouds. The \northrim component gathers elements that have previously been identified as independent features, such as the MBM 18 (L1569) molecular cloud at a distance of $155^{+3}_{-3}$ pc \citep{Zucker2019} and the G203-37 cloud towards (66\degr, -9\degr) that \citet{Snowden95} placed midway through the hot superbubble cavity. Other MBM clouds of interest are listed in Table \ref{tab:Disttable}. We did not push the partition of the \northrim farther because the lines are confused in space and velocity, despite the sparse decomposition of the spectra into individual lines. We reached the same conclusion using another decomposition method called the regularized optimization for hyper-spectral analysis (ROHSA\footnote{https://antoinemarchal.github.io/ROHSA/index.html}). The analysis kindly provided by Antoine Marchal recovered the same structures for the \eastshell and \westrim. This supports our separation. However, it also struggled to identify individual entities in the complex distribution in the northern part. We therefore preferred to keep the \northrim as a single component for our analyses.

Part of the analysed region overlaps the anticentre region studied by \citet{Remy17}, who used a similar method to separate gas clouds. We recovered the same structures in velocity and space in the overlap region: their North Taurus and Cetus components, with minor differences due to their use of data with higher velocity resolution from the Galactic Arecibo L-Band feed array (GALFA) in those directions. Their South Taurus complex mainly contributes to the \northrim in our analysis. Their main Taurus complex only marginally overlaps with our region. The bulk of this cloud lies beyond our analysis boundary and is not discussed here.

    \subsection{Relative motions in the superbubble}
    
The distribution in position and velocity of the \hi line cores that result from the line decomposition allows us to resolve the relative motions that subtend this complex environment. In the original data cubes, the bulk motions are often buried in the overlap of the broad wings of the \hi lines. We display slices through the \hi line distributions for specific velocities in Fig.3. The central velocities of the \hi lines associated with the superbubble range from -15 to +18 \kmpers. This range compares with the [-25, +5] \kmpers range derived by \citet{Reynolds1979} from the \halpha emission. The map at +9 km/s exhibits a coherent circular ring that suggests that we intercept the outer rim of the expanding shell tangentially along these lines of sight. Most of the gas at higher (receding) velocities is indeed seen within the boundary of the outer rim. The slice at +4 km/s shows another circular ring in the south that corresponds to the rim of the \southloop. The expansion velocities of these two rings are in the plane of sky because we see the rims tangentially. The radial line velocities thus indicate that the bulk of the outer rim and of the \southloop recede at +9 and +4 \kmpers , respectively, with respect to us. This means that the \southloop approaches us with respect to the rest frame of the superbubble. The slice at -9 km/s shows the \eastshell, which approaches fast. We give evidence in Section \ref{sec:X-ray} that it moves inside the superbubble. 

When we exclude the internal motion of the \eastshell to follow the outer expansion, the \hi line range is reduced to [-5, +19] \kmpers. With a rest frame velocity of +9 \kmpers, the line data suggest an expansion velocity of about 10-15 \kmpers , which is lower than the 40 \kmpers inferred by \citet{Brown95} from the full extent of the \hi spectra, including the line wings. It is consistent with the 15 to 23 \kmpers velocity found in \halpha \citep{Reynolds1979}. It also compares with the low velocity of 8 \kmpers expected from simulations \citep{Kim17} for a superbubble that expands in a medium with a mid-plane gas density of 1~\percc and a scale height of 104~pc that is powered by a supernova rate of 1 per Myr; this is close to the rate of Orion-Eridanus superbubble \citep{Voss10}. The simulations also show that the compressed swept-up gas behind the shock expands at slightly lower velocities than the shock itself and than the hot phase.

\begin{figure*}
\centering
\resizebox{\hsize}{!}{\includegraphics{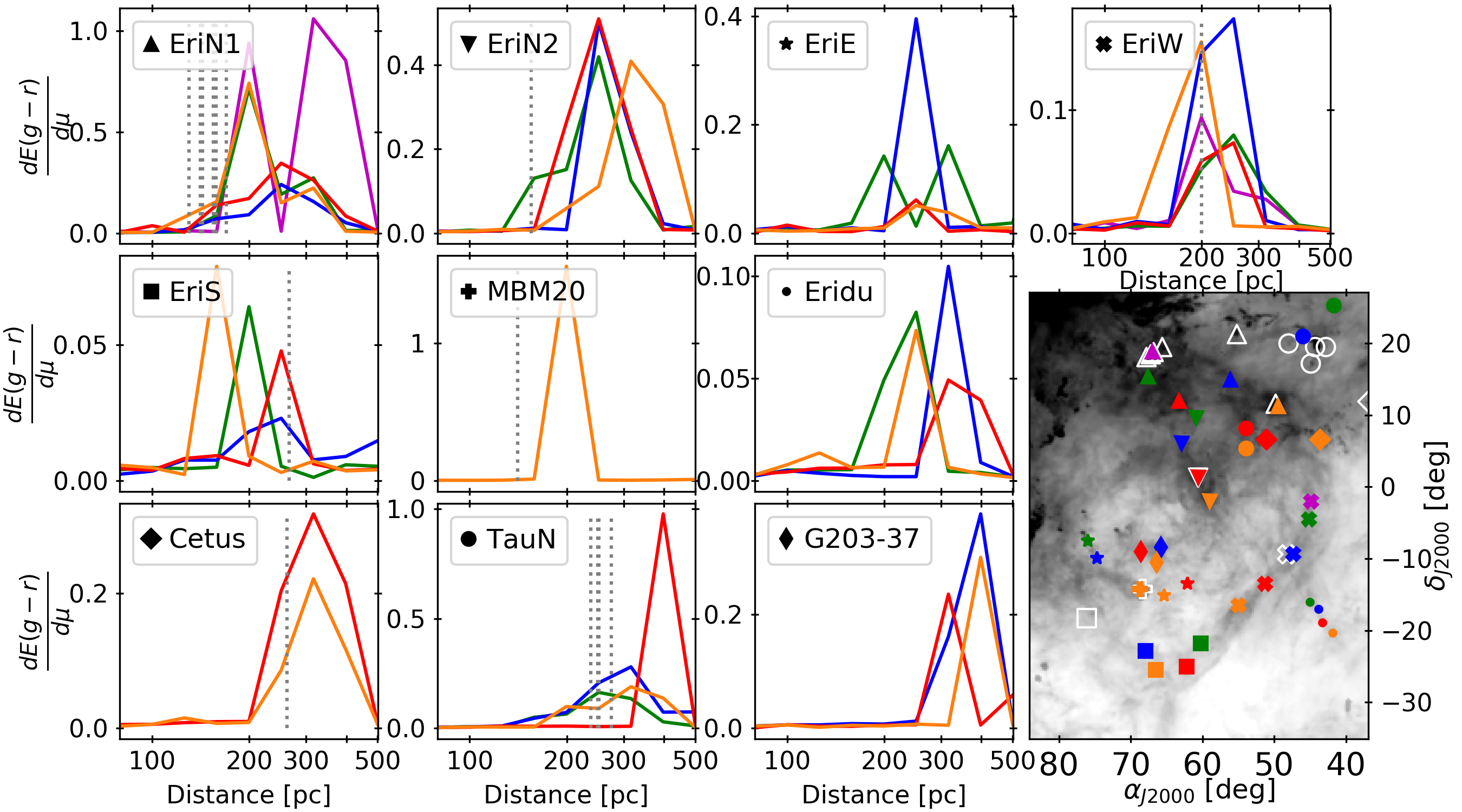}}
\caption{Mean reddening per distance unit profiles towards several directions for each \hi cloud. The mean is computed for each direction over selected profiles in a circle with a radius of 1\degr. The map shows the pointed directions overlaid on the \hi column density. The different markers correspond to the different cloud complexes, and the colours show the different directions for a given cloud. The cloud labels are \northrim (EriN), \westrim (EriW), \eastshell (EriE), \southloop (EriS), and North Taurus (TauN). For clarity, we divided the \northrim into its northern part (EriN1, upwards-pointing triangle), which marks the superbubble boundary, and its southern part (EriN2, downwards-pointing triangle), which overlaps the MBM 18 cloud. The white markers on the map show the MBM clouds. Their shape corresponds to the complex they are associated with. Their distances were derived by \citet{Zucker2019} and are reported in the plots with grey dotted lines.}
\label{fig:DistProf}
\end{figure*}

\begin{table*}[ht]
\caption{Distance and velocity range of the main \hi clouds and the overlapping MBM clouds.}
\begin{tabular}{ l | l l l l l l l }
\label{tab:Disttable}   
 & Velocity & Dist.$^{a}$ & Dist.$^{b}$ & cloud & ($\alpha,\delta$) & MBM Dist.$^{c}$ & MBM Vel.$^{d}$ \\ 
 & km s$^{-1}$ & pc & pc & & J2000 & pc  & km s$^{-1}$ \\ 
 \hline 
 & \multicolumn{7}{c}{superbubble components} \\
 \hline 
\northrim & [4, 25] & [150, 400] & 140$\pm$40 & MBM16 & (49\fdg8, 11\fdg6) & $170^{+2}_{-1}\pm8$ & 7.4 \\ 
 & & & 170$\pm$40 & MBM 17$^{e}$ & (55\fdg2, 21\fdg3) & $130^{+2}_{-1}\pm6$ & 7.9 \\
 & & & & MBM 107 & (66\fdg9, 18\fdg9) & $141^{+4}_{-13}\pm7$ & 8.6 \\
 & & & & MBM 108 & (67\fdg2, 18\fdg5) & $143^{+2}_{-2}\pm7$ & 7.6 \\
 & & & & MBM 18 & (60\fdg6, 1\fdg3) & $155^{+3}_{-3}\pm7$ & 8.0 \\
 & & & & MBM 109 & (67\fdg9, 18\fdg1) & $155^{+10}_{-8}\pm7$ & 6.6 \\
 & & & & MBM 106 & (65\fdg7, 19\fdg5) & $158^{+27}_{-10}\pm7$ & 8.5 \\
\southloop & [-3.8, 7.8] & [150, 300] & & & & \\ 
\eastshell & [-30.8, -2.5] & [200, 300] & & & & \\ 
\westrim & [-2.5, 15.6] &  [150, 250] & 180$\pm$30 & MBM 15 & (48\fdg2, -9\fdg4) & $200^{+27}_{-24}\pm10$ & 4.4 \\
 & & & 210$\pm$40 & & & \\
 \hline 
 & \multicolumn{7}{c}{other complexes} \\
 \hline 
MBM 20 & [0, 4] & [150, 250] & 180$\pm$40 & MBM 20 & (68\fdg8, -14\fdg2) & $141^{+3}_{-2}\pm7$ & 0.3 \\
MBM 22 & & & & MBM 22 & (76\fdg2, -18\fdg3) & $266^{+30}_{-19}\pm13$ & 4.5 \\
Eridu cirrus & [-13, -2] & [200, 400] & 210$\pm$30 & & \\ 
North Taurus & [-6, 4] & [250, 450]  & 370$\pm$50$^{f}$ & MBM 13 & (44\fdg9, 17\fdg2) & $237^{+5}_{-6}\pm11$ & -5.7 \\
& & & & MBM 11 & (42\fdg8, 19\fdg5) & $250^{+5}_{-7}\pm12$ & -6.6 \\
& & & & MBM 12 & (44\fdg2, 19\fdg5) & $252^{+4}_{-6}\pm12$ & -5.3 \\
& & & & MBM 14 & (48\fdg0, 20\fdg0) & $275^{+3}_{-6}\pm13$ & -2.2 \\
Cetus & [-20, -5] & [250, 400] & 370$\pm$50$^{f}$ & MBM 9 & (36\fdg6, 11\fdg9) & $262^{+22}_{-14}\pm13$ & -13.0 \\
G203-37 & [8, 15] & [300, 400] & & &(66\fdg0, -9\fdg0) & & \\
\end{tabular}
\begin{tablenotes}
\item $^{a}$ From the reddening front method described in this paper.
\item $^{b}$ From \citet{Lallement18}.
\item $^{c}$ From \citet{Zucker2019}. The first errors are statistical, and the second give the systematic uncertainties.
\item $^{d}$ From \citet{Magnani1985}.
\item $^{e}$ The association of MBM 17 cloud with \northrim is uncertain because of the velocity confusion in this direction.
\item $^{f}$ It is uncertain whether the association should be with Cetus or North Taurus.
\end{tablenotes}
\end{table*}

        \subsection{Dust reddening distances}
In order to study the 3D gas distribution in the superbubble and to substantiate the cloud separation in velocity, we used the 3D reddening maps of \citet{Green18}. They inferred a reddening profile as a function of distance modulus, $\mu$, for each direction in the sky using the photometric surveys of Pan-STARRS 1 and 2MASS in a Bayesian approach. We located gas concentrations in distance in these profiles by searching for reddening fronts that can be identified as sharp increases in the amount of dust reddening per unit length.
To do so, we took the derivative of their best-fit profiles for each direction in the sky. We display the derivative maps in Fig. \ref{fig:DistMap} for three distances at 200, 320, and 400~pc. The \westrim and MBM 20 clouds are clearly identified around 200~pc, and the \northrim appears between 200 and 320~pc. The CO-bright molecular parts of the \eastshell are visible at 320~pc. The G203-37 cloud appears between 320 and 400~pc.

The 3D partitioning of the dust fronts was not induced by the Bayesian priors that were used to build the reddening profiles. \citet{Green18} have used broad-band photometric measurements of each star and have computed a probability distribution over the stellar distance and foreground dust column. They constrained their prior on the \citet{SFD1998} reddening map, which is a scaled version of an optical depth map of the dust thermal emission integrated along the lines of sight, without information along the radial dimension. The spatial correlation we find between the dust maps in distance and the morphology of the clouds we isolated strengthens our confidence that the different complexes are separated in velocity as well as in distance.

We constrained the distance range to each cloud complex using the reddening front information towards several directions in each complex. Their locations are marked in Fig. \ref{fig:DistProf}. For each direction, we selected derivative profiles within a radius of 1\degr\ , and we averaged those whose peak $\frac{dE(g-r)}{d\mu}$ value exceeded 80\% of the maximum. The results are shown in Fig. \ref{fig:DistProf}. Distance ranges based on these profiles are given in Table \ref{tab:Disttable} together with the velocity ranges observed in the \hi data. Table \ref{tab:Disttable} also lists distances to MBM molecular clouds that appear to be associated in space and velocity with our cloud complexes. Their distances have been derived by \citep{Zucker2019} using the same reddening method, but choosing individual stars. \citet{Lallement18} constructed another 3D dust map using inversion methods and the stellar surveys of Gaia, 2MASS, and the APO Galactic Evolution Experiment (APOGEE) DR14. Three vertical cuts across the Galactic disc towards the Galactic longitudes of 192\degr, 205\degr\, , and 212\degr\ yield distances to our gas complexes that are listed in Table \ref{tab:Disttable}. 

The different measurements in Table \ref{tab:Disttable} indicate that most of the clouds associated with the superbubble, such as the \northrim and \westrim, the \eastshell, \southloop, and MBM 20, lie at distances ranging between 150 and 250 pc. The distances and velocities we obtain for the different complexes are consistent with those found with NaI and CaII interstellar absorption lines towards early-type stars throughout the region \citep{Welsh05}. We note, however, that distances based on the reddening information from \citet{Green18} tend to yield higher values than towards individual stars. This is partly due to their binning in space and in $\log(\mu),$  which was adapted to large-scale 3D mapping in the Galaxy, and to the lack of stars at high Galactic latitudes. The lower bound of our distance estimates should also be considered with care as they approach the shortest range allowing large enough stellar samples to infer the average reddening and to recognise the shape of the cloud. Adding stars from Gaia alleviates the biases. All sets of values nevertheless give weight to a superbubble orientation with its close end towards Eridanus. 

A series of MBM clouds (MBM 16, 106, 107, 108, and 109) is associated with the \northrim and more specifically with the outer rim because of their location and velocity. They gather at distances between 140 and 170~pc. The MBM 18 cloud, at a comparable distance, is also associated with the \northrim in \hi and CO. Together they suggest that the broad \northrim complex combines gas from the outer rim seen edge-on and face-on (see the middle right panel of Fig. \ref{fig:HI9kms}). We therefore conclude from the synergy in distance and velocity of the \northrim, \southloop, and \westrim that they represent the front face of the superbubble with respect to us.

Figure \ref{fig:DistMap} also shows the G203-37 cloud at a distance between 300 and 400~pc, in agreement with \citet{Snowden95}. These authors had placed it midway through the hot superbubble cavity.

The MBM 22 cloud is seen in a direction close to the rim of the \southloop, and its velocity is consistent with the \southloop range, but its distance of $266^{+30}_{-20}$~pc places it behind the \southloop. This is corroborated by the maps in Fig. \ref{fig:DistMap}, which show that the \southloop is most conspicuous in the left (200~pc) panel and fades out at 320~pc, whereas MBM 22 appears at 320~pc and fades out at a larger distance.
MBM 9 and the clouds of Cetus and North Taurus also lie outside and behind the superbubble, at distances between 250 and 400~pc. We note that the Cetus and North Taurus clouds are part of larger complexes that extend well beyond the western side of the analysed region \citep{Remy17}.

\begin{figure}
\centering
\resizebox{0.9\hsize}{!}{\includegraphics{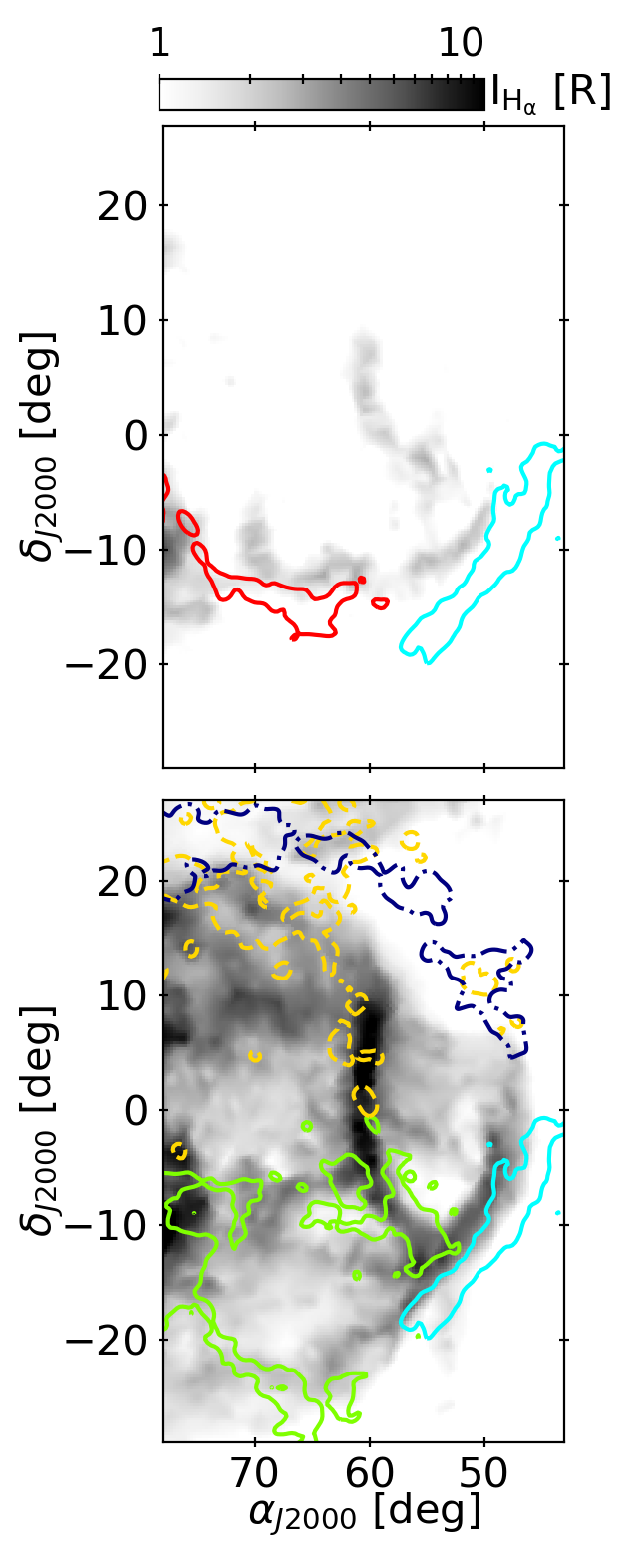}} % 0.5 for referee
\caption{\halpha line intensity maps for two velocity ranges, [-60, -11] (\textit{top}) and [-11, 50] \kmpers (\textit{bottom}), overlaid with \nhi column density and \wco emission contours outlining the different cloud components: the \eastshell in \hi (red, $1.5 \times 10^{20}$~cm$^{-2}$), the \westrim in \hi (cyan, $4 \times 10^{20}$~cm$^{-2}$), the \southloop in \hi (green, $2.5 \times 10^{20}$~cm$^{-2}$), and the \northrim in CO (yellow dashed line, 1 \wcounit). The \nhi and \wco maps are smoothed with a Gaussian kernel of 0\fdg28 to display the contours. A specific \northrim contour is shown in \hi for central line velocities of +9 \kmpers (dark blue dash-dotted line, 18 K). It was smoothed with a Gaussian kernel of 1\fdg5 before the contours were computed.}
\label{fig:HaMap}
\end{figure}

    \subsection{MBM 20 and the edge of the superbubble}
 
Inspection of relative distances in Figs. \ref{fig:DistMap} and \ref{fig:DistProf} suggests that MBM 20 and the closest clouds of the \southloop may be interacting with the Local Bubble at about 150~pc in the picture of \citet{Burrows93}. This is consistent with the velocity configuration discussed above, where the \southloop approaches us compared to the rest of the superbubble. 

A clear \hi and CO component in our separation corresponds to the well-studied MBM 20 (L1642) cloud \citep{Russeil2003}. It is clearly visible in the 200~pc map of Fig. \ref{fig:DistMap} at $\alpha = 68\fdg8$ and $\delta = -14\fdg2$. Its distance of $141^{+3}_{-2}$ pc \citep{Zucker2019} or 180$\pm$40~pc \citep{Lallement18} places it near the boundary between the superbubble and the Local Bubble. \citet{Snowden95} used X-ray absorption to place MBM 20 inside the Local Bubble, thus pushing the closest superbubble wall beyond 140~pc. The cometary tail of MBM 20 is visible in \hi. It points to the north-east of the compact CO cloud (see Fig. \ref{fig:DustModel}) and extends over 10~pc (at 140~pc). It presents a velocity gradient whose tail tip moves faster away from us than the molecular head, in agreement with having been swept up by the expanding medium of the Local Bubble rather than by the \southloop. MBM 20 may therefore lie just in front of the edge of the superbubble. We note that despite similar orientations and some overlap in direction with the \eastshell, the two clouds are distinct and have different velocities, as is shown in Table \ref{tab:Disttable}.

        \subsection{Cloud relations to the \halpha filaments}
        
The \hi and CO structures we isolated can be compared to the \halpha intensity map. \halpha line emission at 656.3~nm arises from the recombination of an electron onto ionised hydrogen, so that the lines trace the ionised gas at the interface with colder atomic hydrogen, where the ionisation fraction is still high, but the density is also high enough for recombination to occur. Figure \ref{fig:HaMap} displays \halpha line intensity maps integrated over [-60, -11] and [-11, 50] \kmpers. We overlaid \hi contours of specific clouds. The 12 \kmpers resolution of the WHAM \halpha data prevents detailed velocity comparisons between the \hi, CO, and \halpha lines, but it can still capture interesting relations. As the neutral \hi structures surround the recombining shells with a small angular offset, Figure \ref{fig:HaMap} outlines a mosaic of boundaries between hot sub-bubble interiors and their colder outer shells, which are compressed by the hot-gas expansion \citep[see Figs. 3 and 4 of][]{Krause13}.

The \eastshell is related to a specific \halpha filament. Both structures move inside the superbubble towards us (negative velocities) at a distance constrained by the dust to be about 300~pc. The southern part of the \southloop might be associated with the \halpha filament called Arc C \citep{Johnson78}. The round structure of the loop in \hi and \halpha and its radius of $\sim$35~pc at a distance of 200~pc are reminiscent of an old supernova remnant, but they may also reflect global oscillations of the hot gas in the superbubble. Such oscillations are induced by off-centred supernova explosions in a hot cavity \citep{Krause14}. We further discuss the origin of the loop in the next sections. The \westrim is related to the Arc B filament in \halpha \citep{Johnson78}. We see an elongated shock wall almost edge-on, with velocity excursions of -10 to +7 \kmpers about the rest velocity of +9 \kmpers. The closest part approaches us and is superimposed on receding parts that lie farther away. The \halpha emission from the rest of the outer rim is likely absorbed by the high gas column densities pertaining to the \northrim (typical visual extinctions of 0.4-1.8 magnitudes up to a distance of 300~pc along the northern part of the outer rim; this is outlined in dotted contours in Fig. \ref{fig:HaMap}. Towards the \westrim, the visual extinctions range from 0.1 to 0.4 magnitudes).

\begin{figure*}
\centering
\includegraphics[width=17cm]{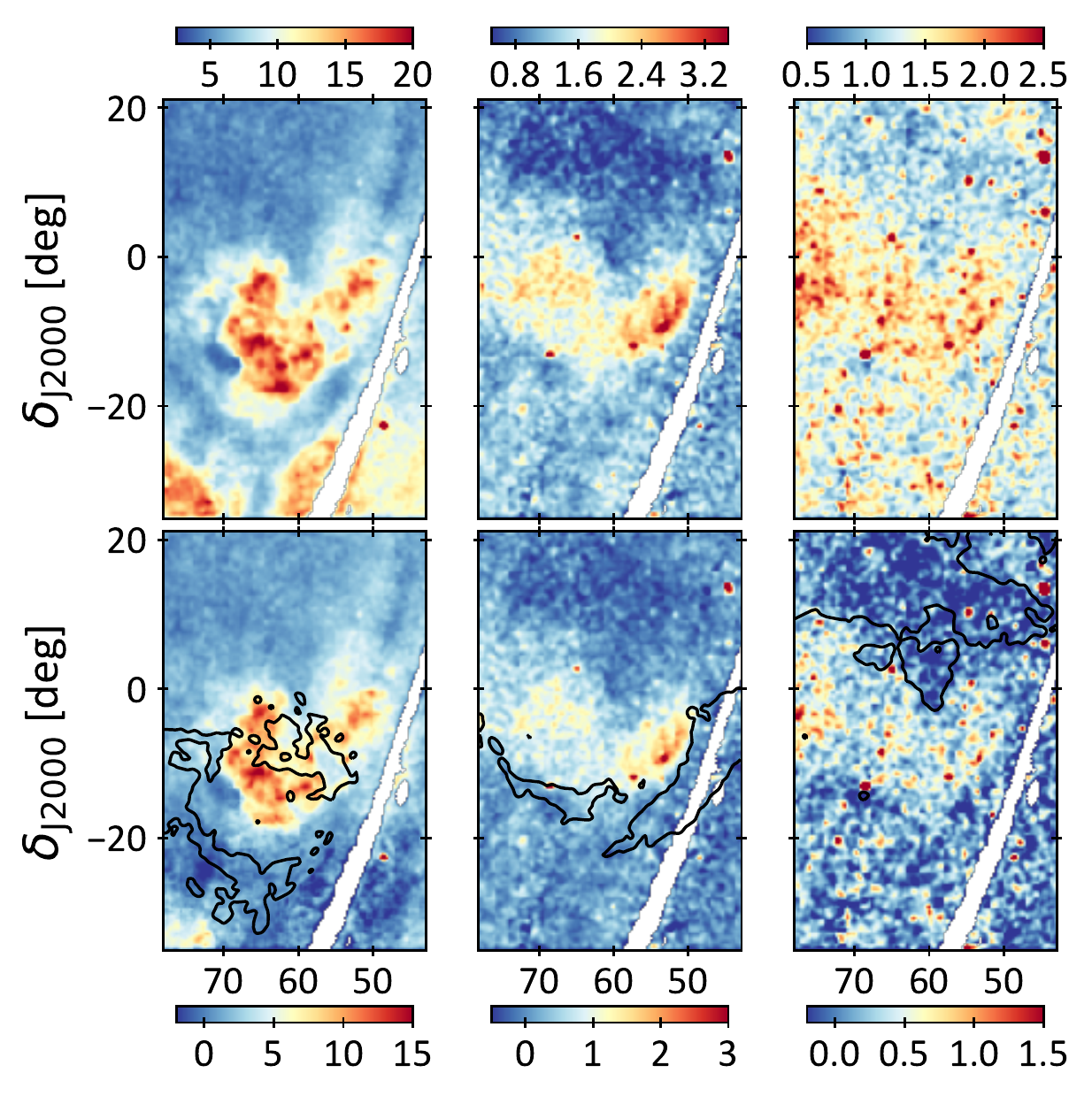}
\caption{\ROSAT intensity maps in units of $\rm{10^3\ counts\ s^{-1}\ sr^{-1}}$ in the 0.25 keV band (covering between 0.11 and 0.284~keV, \textit{left column}), in the 0.75 keV band (covering between 0.44 and 1.21 keV, \textit{middle column}), and in the 1.5 keV band (covering between 0.73 and 2.04 keV \textit{right column}) from the all-sky survey in the \textit{top row}, and the same maps corrected for the Local Bubble foreground emission and for a halo background emission in the \textit{bottom row} (according to equation \ref{eq:IEri}). \hi column density contours outline different cloud complexes: the \northrim absorbs X-rays at all energies and is displayed in the 1.5 keV map (at \nhi = 1.5 $\times 10^{21}$~cm$^{-2}$). The \westrim cloud bounds the X-ray emission at all energies and is displayed in the 0.75 keV map (at \nhi $=2 \times 10^{20}$~cm$^{-2}$). The \eastshell cloud bounds the southern parts of the 0.75 and 1.5 keV emissions, but does not absorb the 0.25 keV emission. It is displayed in the 0.75 keV map (at \nhi = 1.4 $\times 10^{20}$). Additional absorption at 0.25 keV relates to the \southloop (\nhi = 2 $\times 10^{20}$~cm$^{-2}$) and to MBM 20. The latter is not displayed, but is visible as the dark blue spot around $\alpha=68\fdg8$ and $\delta=-14\fdg2$ in the top 0.25 keV map. For clarity, we smoothed he X-ray maps and \hi contours with a Gaussian kernel of 0\fdg3 for display and maksed underexposed regions in the \ROSAT survey in white.}
\label{fig:ROSATMap}
\end{figure*}

The brightest \halpha arc, Arc A, extends vertically in the centre of the positive-velocity map in Fig. \ref{fig:HaMap}. Its relation to  the superbubble is still debated; see \citet{Pon14} for a discussion that takes the proper motion and radial velocities, as well as X-ray and \halpha emission and absorption studies, into account. Based on the width and intensities of the \halpha lines, these authors argued that the \halpha filaments are either elongated sheets seen edge-on or objects that are out of equilibrium as a result of ionised clouds that are compressed by shocks that currently recombine. While Arc B seems to comply with the edge-on sheet hypothesis, the situation is not clear for Arc A because of the confusion along the line of sight, the spread of \halpha and \hi emission over a wide range of velocities, and absorption from atomic and dark neutral gas (see Paper II) and from molecular gas in MBM 18. As reported by \citet{Pon14}, the coincidence in space and velocity between Arc A and MBM 18 tends to favour an association of the arc with the superbubble. Figure \ref{fig:HaMap} shows that the northern part of Arc A nicely follows an arc of molecular clouds belonging to the \northrim component and that Arc A is partially absorbed by the dense gas in MBM 18 near $\alpha_{J2000}=60\fdg6, \delta_{J2000}=1\fdg3$. The dust associated with the molecular arc is visible at close distance in Fig. \ref{fig:DistMap}, but not at 400~pc, nor beyond. The distance to Arc A is thus constrained to be between that of MBM 18 \citep[$155^{+3}_{-3}$~pc,][]{Zucker2019} and an upper bound of about 400~pc. Arc A therefore likely represents an inner shell of the superbubble seen edge-on, as for the \eastshell. 

\section{X-ray emission}
\label{sec:X-ray}

The Eridanus X-ray enhancement arising from diffuse hot gas has been extensively studied to constrain the properties of the hot superbubble interior and the distribution of foreground gas seen in absorption against the bright X-rays \citep{Burrows93, Snowden95, Heiles99}. We now compare the spatial distribution of the different clouds we isolated with that of the X-rays recorded in three energy bands with \ROSAT, and we take advantage of our study of the total gas column densities in the atomic, dark, and molecular gas phases (see Paper II) to quantify the X-ray absorption.

Figure \ref{fig:ROSATMap} displays the \ROSAT maps recorded in the 0.11-0.284~keV (R1+R2), 0.44-1.21 keV (R4+R5), and 0.73-2.04 keV (R6+R7) energy bands \citep{Snowden94}. We refer to these bands below as the 0.25, 0.75, and 1.5 keV bands, respectively. Two main features have originally been identified \citep{Burrows93}. The first, called Eridanus X-ray enhancement 1 (EXE1), appears in the 0.75 and 1.5 keV maps as a crescent bounded to the west by the \westrim clouds, to the south by the \eastshell, and to the east, it extends towards Orion. Its northern extent is unknown because it is obviously heavily absorbed, even at 1.5 keV, by the 1.4 $\times 10^5$ solar masses of gas associated with the \northrim (see Fig. \ref{fig:ROSATMap}). EXE1 was associated with the interior of the superbubble by \citet{Burrows93}. The brightest part of the crescent at 0.75 keV is located between the \halpha arcs A and B. This region was found to be a small independent expanding shell in \halpha by \citet{Heiles99}. 

The second main feature (EXE2) is visible only at 0.25 keV, below $\delta=-10$\degr. This soft excess is obviously barred by absorption from the gas in the \southloop and by the compact MBM 20 cloud, which creates a marked dip in X-rays around $\alpha=68\fdg8$ and $\delta=-14\fdg2$. The \eastshell, however, leaves no obvious absorption imprint. The bright part of EXE2 that partially overlaps the \southloop was first interpreted as hot gas originating from a supernova remnant or a stellar wind bubble \citep{Burrows93}, but the absence of an OB star in this direction led \citet{Heiles99} to propose that hot gas leaks out of EXE1 at the rear of the superbubble through a break in its shell and that the gas cools in the external ISM. Because the soft emission extends below $-25$\degr \ in declination, \citet{Snowden95} instead interpreted EXE2 as soft background emission arising from the million-degree gas in the Galactic halo and being shaped into a roundish feature by intervening gas absorption. We revisit the properties of EXE1 and EXE2 below.

\begin{figure}
\resizebox{\hsize}{!}{\includegraphics{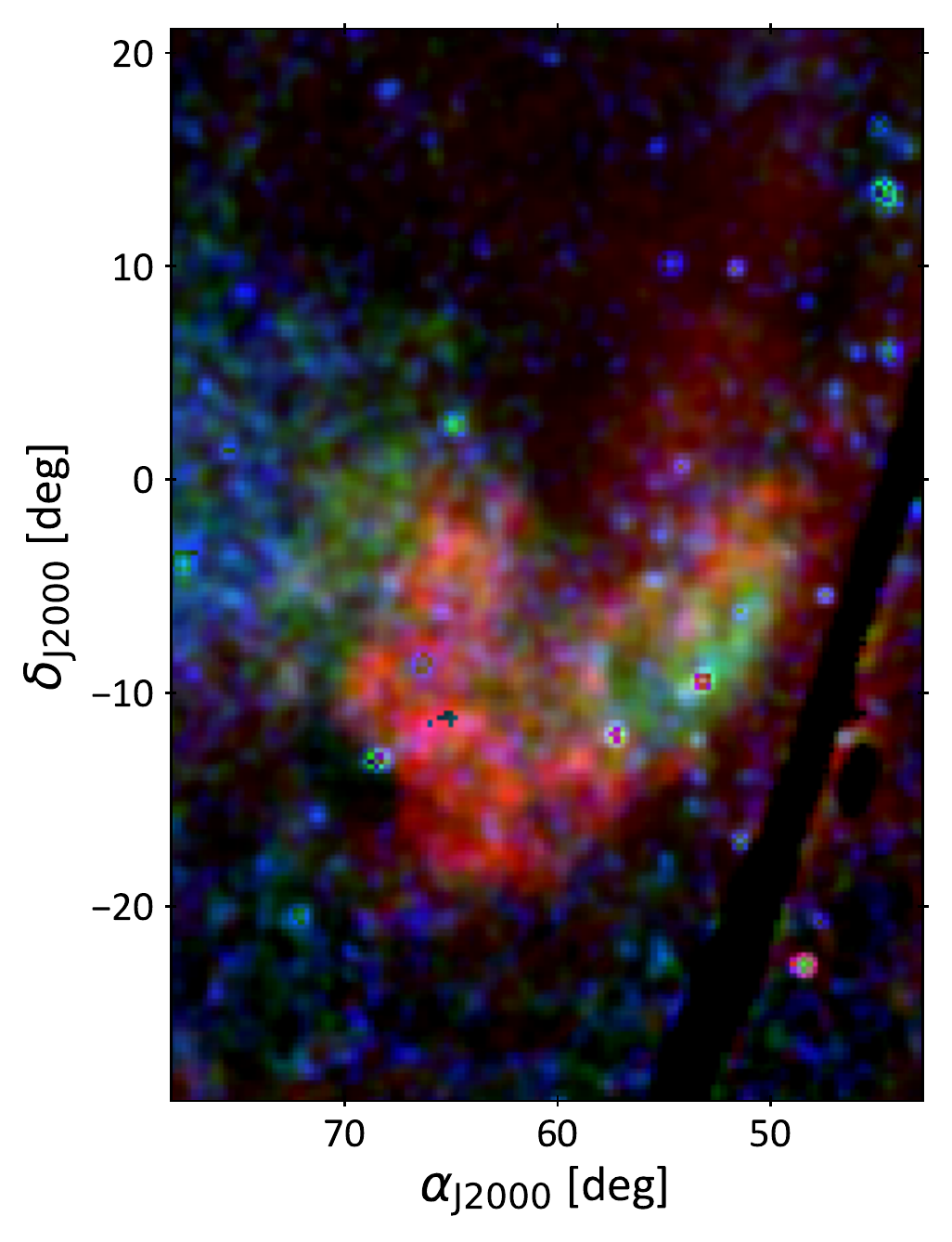}}
\caption{Composite image of the \ROSAT intensities obtained at 0.25 keV (red), 0.75 keV (green), and 1.5 keV (blue) after subtraction of the Local Bubble foreground emission and of the halo background emission (according to equation \ref{eq:IEri}). The underexposed regions in the \ROSAT survey are masked in black.}
\label{fig:RGBXMap}
\end{figure}

        \subsection{Subtracting foreground and background emission}
        \label{sub:EXE2}

We followed the simple slab model of \citet{Snowden95} and modelled the observed X-ray intensity in each energy band and direction as
\begin{equation}
\rm{I_{obs}(\alpha,\delta) = I_{Local\ Bubble} + I_{Eri}(\alpha,\delta) \, e^{-\tau_X(\alpha,\delta)} + I_{halo}\, e^{-\tau_{tot\,X}(\alpha,\delta)} }
,\end{equation}
\label{eq:Imodel}
where $\rm{I_{Local\ Bubble},\ I_{Eri}\ , and\ I_{halo}}$ represent the X-ray intensities arising from the Local Bubble, the Orion-Eridanus superbubble, and the halo, respectively. The foreground and background emissions from the Local Bubble and the halo are assumed to be uniform. $\tau_X$ denotes the X-ray optical depth of the absorbing gas located in front of the superbubble, and $\tau_{tot\,X}$ denotes the total optical depth along the line of sight. 

In order to compute the X-ray optical depths, we used the \nh gas column densities inferred in the different gas phases and different cloud complexes from our combined \hi, CO, dust, and \g-ray study, following the same method to map and quantify the amount of gas as applied earlier in other nearby regions \citep[see Paper II,][]{PlanckXXVIII15, Remy17}. \citet{Snowden95} have instead used the dust emission map at 100 $\mu$m to trace \nh, but they did not correct the map for biases that have since then been found to be significant (e.g. spatial variations in dust temperature and spectral index \citep{PlanckXI14} and the non-linear increase of the dust emissivity per gas nucleon with increasing \nh \citep{PlanckXI14, PlanckXXVIII15, Remy17}. We derived the optical depths from our column density maps using
\begin{equation}
\tau_X(E_{\rm X}) = \sigma(E_{\rm X}, N_{\rm H}) \times N_{\rm H}
\label{eq:tauX}
,\end{equation}
with $E_{\rm X}$ the X-ray band and $\sigma$ the energy-band-averaged photoelectric absorption cross section from \citet{Snowden94}. For the 0.25 and 0.75 keV bands, we took an incident thermal spectrum with a temperature of 1.3 MK close to the mean colour temperature observed towards the Eridanus X-ray excesses (see below). For the 1.5 keV band, we assumed a power-law incident spectrum with an index of -2. We verified that the choice of temperature and spectral index around these values has a small effect on the result. The optical depth maps obtained for the total gas column densities $\tau_{tot\,X}$ are presented in Fig. \ref{fig:tauX} for the three energy bands.

We find that the \northrim is optically thick in the lower energy bands and is still partially thick ($0.5 \lesssim \tau_{tot\,X} \lesssim 1$) at 1.5 keV, which explains the northern shadows in all plots of Fig. \ref{fig:ROSATMap}. Inspection of the optical depth maps for individual cloud complexes indicates that the \westrim is thick at 0.25 keV, nearly so at 0.75 keV, and mostly thin ($0.1 \lesssim \tau_X \lesssim 0.2$) at high energy. The \southloop is thick at low energy and mostly thin in the two upper bands. The \eastshell is thick at low energy ($1 \lesssim \tau_X \lesssim 2.5$), partially thick at 0.75 keV ($0.1 \lesssim \tau_X \lesssim 0.4$), and thin at high energy. The molecular head of MBM 20 is optically thick at low energy, nearly so at medium energy, but partially transparent at high energy.

In order to estimate upper limits to the emission from the Local Bubble in the three \ROSAT bands, we studied the distribution of X-ray intensities versus gas column density towards a set of heavily absorbed and nearby clouds: towards MBM 20 (which is likely located at the interface between the Local Bubble and the Eridanus superbubble), MBM 16, MBM 109, and MBM 18. To limit the effect of noise fluctuations towards these faint regions, we calculated the minimum intensities averaged over the high-\nhtot pixels in each region, at \nh $> 7 \times 10^{20}$~cm$^{-2}$ in the low band and \nh $> 1.5 \times 10^{21}$~cm$^{-2}$ in the upper bands, and we excluded pixels with X-ray point sources. We find an upper limit to the Local Bubble intensity of $4.0 \times 10^3$~cts s$^{-1}$ sr$^{-1}$ at 0.25 keV, which coincides with the estimate of \citet{Snowden95}. We find upper limits of $0.8 \times 10^3$~cts s$^{-1}$ sr$^{-1}$ at 0.75 keV and $1.1 \times 10^3$~cts s$^{-1}$ sr$^{-1}$ at 1.5 keV, which are consistent towards the four clouds. 

In order to estimate the background halo intensities in directions away from the superbubble, we corrected the observed X-ray maps for the Local Bubble emission and for the total gas absorption using $\rm{I_{halo} = (I_{obs}-I_{Local\ Bubble})\,e^{+\tau_{tot\,X}}}$. We averaged the intensities over the low-absorption region at $50^{\circ} \leq \alpha \leq 60^{\circ}$ and $\delta \leq -25^{\circ}$ to avoid the underexposed survey stripes. We obtained intensities of $22.0 \times 10^3$~cts s$^{-1}$ sr$^{-1}$ at 0.25 keV, $0.5 \times 10^3$~cts s$^{-1}$ sr$^{-1}$ at 0.75 keV, and $0.2 \times 10^3$~cts s$^{-1}$ sr$^{-1}$ at 1.5 keV. 

The bottom row of Fig. \ref{fig:ROSATMap} shows the residual intensities obtained in the three \ROSAT bands after the uniform Local Bubble emission and the absorbed halo background were removed from the original data, 
\begin{equation}
I_{\rm Eri} \, e^{-\tau_X} = I_{obs} - I_{Local\ Bubble} - I_{halo} \, e^{-\tau_{tot\,X}}. 
\label{eq:IEri}
\end{equation}
These maps show the distributions of X-rays that are likely produced by the hot gas that pervades the superbubble, but the spatial distributions are still modulated by photoelectric absorption by internal and/or foreground clouds. 
The residual intensities in the two upper bands are consistent with zero across the reference region we used to estimate the halo background. As our estimate of the halo intensity at 0.25 keV is 45\% lower than that of \citet{Snowden95}, the data are not as over-subtracted at low declinations as in their analysis. Only a few patches of small negative residuals remain at $\delta < -20^{\circ}$. They may be due to anisotropies in the Local Bubble and/or in the halo at low X-ray energies. 

Fig. \ref{fig:RGBXMap} presents a colour image of the superbubble emission that remains after the Local Bubble foreground and halo background are subtracted according to equation \ref{eq:IEri}. It reveals several colour gradients in the X-ray emission: a marked hardening to the east towards Orion, and a more moderate hardening in the expanding region that is enclosed between the \halpha arcs A and B. As pictured by \citet{Ochsendorf15}, the hardening towards Orion traces hotter gas that is energised near Orion and flows into the colder Eridanus region. This hot gas may come from the supernova that is responsible for Barnard's Loop, the eastern side of which is visible in \halpha, but the western side of which has long been in the process of merging with the superbubble hot gas.

The two bright emission regions seen at $\delta<-20^{\circ}$ in the original 0.25 keV map (top left plot of Fig. \ref{fig:ROSATMap}) are well accounted for by the Local Bubble and halo intensities, whereas significant emission remains visible inside the \southloop after the foreground and background emissions are removed. In addition, the \eastshell that bounds the hard EXE1 emission leaves no absorption shadow at 0.25 keV despite its strong optical depth in this band. Together, these facts suggest that hot gas inside the superbubble spans distances beyond the \eastshell in hard X-rays and in front of the \eastshell towards the \southloop in soft X-rays. By correcting the soft "Eridanus-born" X-ray intensity (bottom left plot of Fig. \ref{fig:ROSATMap}) for the amount of absorption expected from \southloop gas, in other words, by plotting 
$(I_{obs} - I_{Local\ Bubble} - I_{halo} \, e^{-\tau_{tot\,X}}) / e^{-\tau_{X_{South}}}$ , where $\tau_{X_{South}}$ is the optical depth of the \southloop, we verified that the X-ray gap that separates the \southloop from the hard emission shining between arcs A and B can be fully accounted for by absorption in the \southloop rim. The in situ emission may thus be continuous.

\begin{figure}
\resizebox{\hsize}{!}{\includegraphics{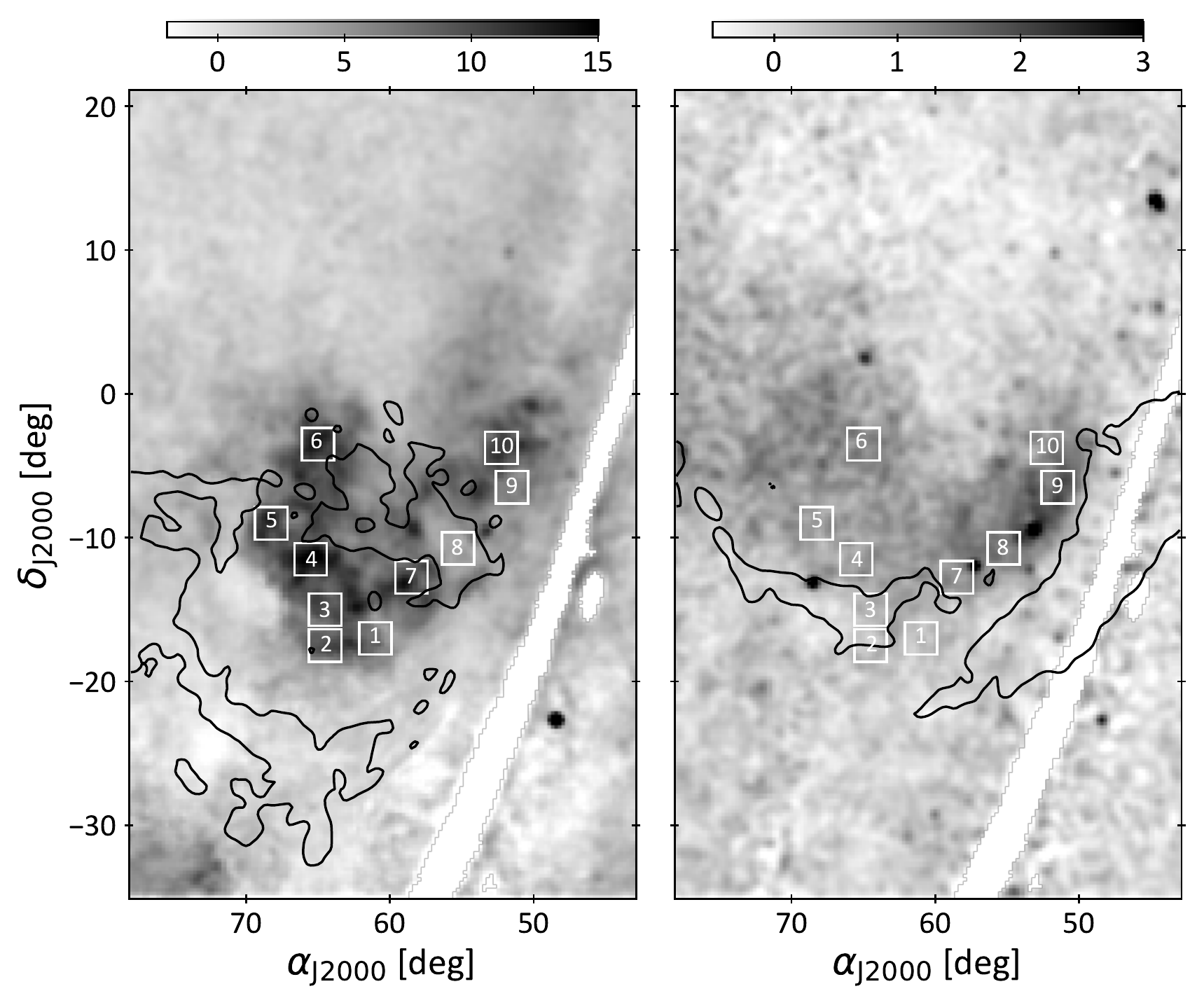}}
\caption{Locations of the regions chosen for X-ray spectral modelling compared with the directions of the \westrim, \eastshell, and \southloop cloud complexes (same contours as in Fig. \ref{fig:ROSATMap}), displayed over the intensities obtained in the 0.25 (\textit{left}) and 0.75 (\textit{right}) keV bands after correcting for the Local Bubble and halo emissions (same as the bottom row of Fig. \ref{fig:ROSATMap}). The underexposed regions in the \ROSAT survey are masked in white.}
\label{fig:Xmapech}
\end{figure}

\begin{figure*}
\includegraphics[width=17cm]{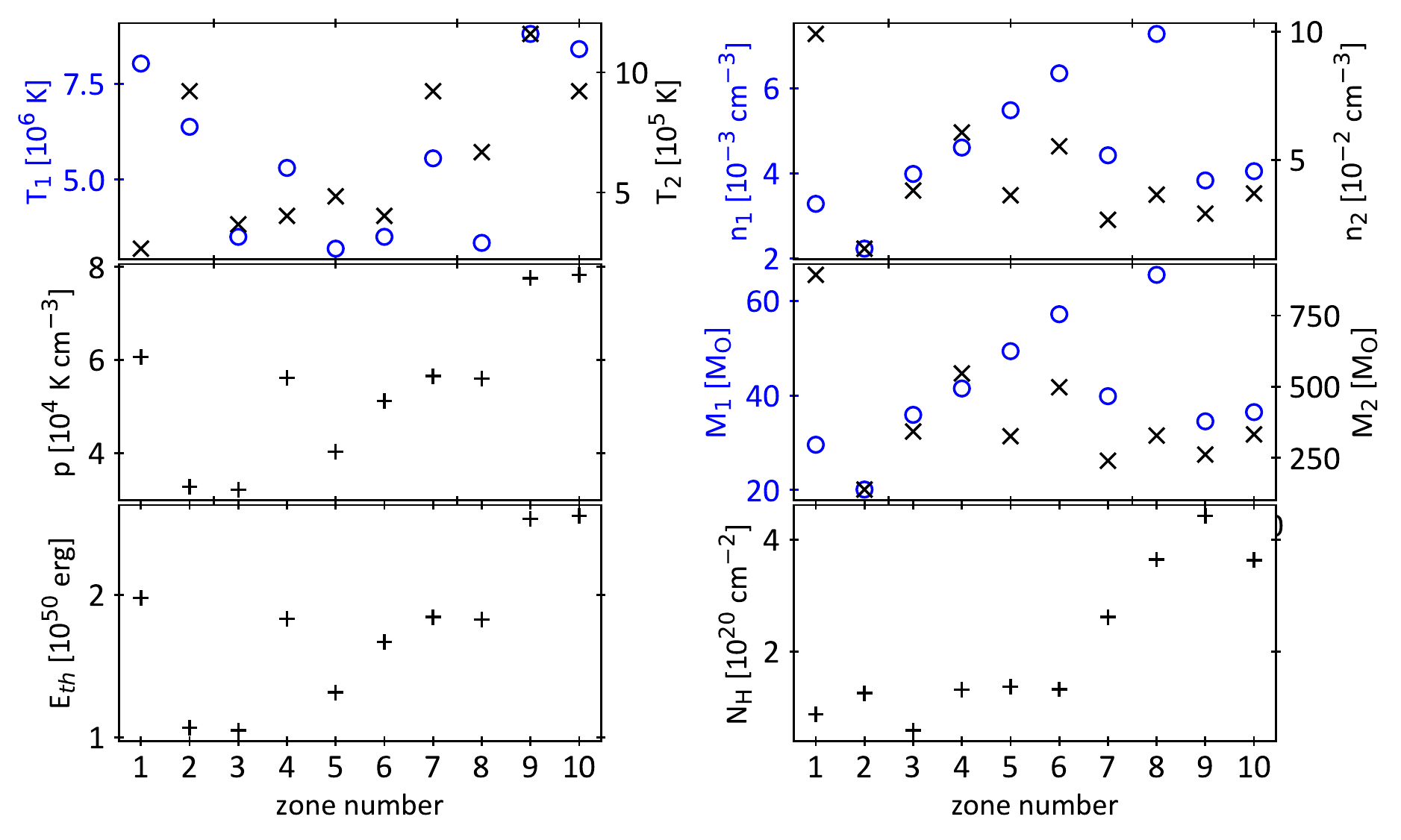}
\caption{Temperatures, gas densities and masses, and common pressure obtained from the superposition of two hot plasma bubbles (mekal models, see Sect. \ref{sub:Xprop}), 80~pc in diameter, towards each of the areas sampled in Fig. \ref{fig:Xmapech}. Blue circles (left scale) and black crosses (right scale) mark the hotter and colder media, respectively. The lower right plot gives the mean \nh column density we used for absorption in each area.}
\label{fig:Xprop}
\end{figure*}

    \subsection{Hot-gas properties}
    \label{sub:Xprop}
    
We studied the average properties of the hot X-ray emitting gas in a sample of ten directions spanning different parts of the EXE1 and \southloop regions, as displayed in Fig. \ref{fig:Xmapech}. The 2\fdg25 width of each square area in the sample is close to the field of view of the ROSAT position sensitive proportional counter (PSPC) detector, whose diameter is 2\degr. For each area, we calculated the mean X-ray intensity that remains in the three energy bands after the Local Bubble foreground and halo background are subtracted (according to equation \ref{eq:IEri}). We considered two hardness ratios: the 0.25 to 0.75 keV and the 0.75 to 1.5 keV intensity ratios. We also calculated the mean gas column density, \nh, in each area, using specific cloud components: only the nearby \southloop for the first three areas that sample the soft EXE2 emission without an indication of strong absorption by the optically thick \eastshell; using the \southloop and the \eastshell for regions 4 to 8 because they sample the hard EXE1 emission behind the \eastshell; and using the whole gas for regions 9 and 10 because they might intercept the faint edge of the \northrim. The resulting column densities are given in Fig. \ref{fig:Xprop}. We used XSPEC v12.9 \citep{Arnaud1996} and the ROSAT PSPC C response (appropriate for the survey) to model the expected flux in the three energy bands from the Mewe-Kaastra-Liedahl (mekal) model of thermal emission arising from a tenuous hot plasma with solar abundances. The mekal model includes bremsstrahlung radiation and the relevant atomic shell physics for line emissions \citep{Mewe85,Liedahl95}. We selected the T\"{u}bingen-Boulder (tbabs) model of interstellar absorption and scanned temperatures from 0.01 to 10 keV in log steps of 0.1. The absorption column was fixed to the mean value in each area. The free parameters of the model are thus the gas temperature, $T$, and the plasma emission measure (i.e. the volume integral $\int n_e\,n_{\rm H}\,dV$ of the hydrogen volume density, $n_{\rm H}$, times the electron volume density, $n_e$, the latter becoming 1.2 $\int n_{\rm H}^2\, dV$ for a fully ionised plasma with 10\% helium by number).

We considered a single-temperature model and a mixed model with two temperatures. In the former case, hardness variations across the field would primarily be due to absorption features. The latter case would imply two pockets of hot gas at different temperatures that partially overlap in direction, but are separated in distance and position relative to the \eastshell. The model can accommodate more complex interleaved geometries along the lines of sight as long as the absorbing \nh slab stands in front of both hot gases. We first compared the modelled and observed hardness ratios in each area and found that a single temperature cannot account for the three-band colour distribution in any of the ten areas. The temperatures inferred from the 0.25 to 0.75 keV intensity ratios vary from 1.1 to 1.5 MK around a mean of 1.3 MK, which corroborates our choice of temperature for the derivation of optical depth maps in the former section and agrees with the former estimates of \citet{Guo95}. In particular, the temperature of 1.5 MK we find towards field 7 is consistent with the values of $1.75^{+0.35}_{-0.34}$ and $1.64^{+0.42}_{-0.17}$ found by \citet{Guo95} in directions enclosed in this field. In all fields, the inferred temperatures do not predict enough emission in the 1.5 keV band, however.

In order to solve the two-temperature model in each area (which requires four parameters against three data points), we furthermore assumed that the two plasma pockets along the line of sight are in pressure equilibrium, in agreement with simulations \citep{Kim17}. This is substantiated by the 0.7 Myr sound crossing time of the 80~pc wide region that emits X-rays (for 1.3 MK). We adopted a typical 80~pc size for the pockets in view of the 70~pc diameter of the \southloop at a distance of 200~pc and of the apparent EXE1 size of 100~pc at the further distance of $\sim$250~pc of the \eastshell.
The temperature, gas density, and mass we found for the two plasmas in each area are displayed in Fig. \ref{fig:Xprop}, together with their common pressure and the total thermal energy content of the two plasmas. The mass and energy estimates assume the same spherical geometry, 80~pc in diameter, for the two plasmas. We did not attempt to propagate the original count rate uncertainties to the results shown in Fig. \ref{fig:Xprop} because of the striped calibration problems across the field, the simplicity of a uniform foreground and background subtraction, and the simplicity of a single slab that absorbs the emission from the two plasmas (as opposed to a more complex distribution of the material in front, in between, and inside the emitting regions). The model uncertainties dominate those in the X-ray data.  

We nevertheless find a limited dispersion in the properties of the two plasmas across the different areas. Temperatures range from 3 to 9 MK in the hotter plasma and between 0.3 and 1.2 MK in the colder plasma. They compare well with the asymptotic values of 1.6-2.1 MK that were found in simulations for a superbubble resembling that of Orion-Eridanus \citep[model n1-t1 of][]{Kim17}.
The gas enclosed between the \halpha arcs A and B (areas 8 to 10) appears to be globally hotter and at a higher pressure than in the rest of the region, which supports the \citet{Heiles99} view of a younger and faster-expanding zone between the arcs. The remaining data points are roughly consistent with two 80 pc wide bubbles that encompass all of them. 

The low plasma masses we find indicate that most of the hot gas inside the superbubble has already cooled down radiatively. According to simulations \citep{Krause14}, soft X-rays trace the current thermal energy content of a superbubble, which we find to be about $(1-2.4) \times 10^{50}$ ergs in Eridanus. They do not trace the cumulative energy injected during its lifetime. The thermal energy in the hot gas represents only a few percent of the current kinetic energy of the expanding \hi gas \citep[$3.7 \times 10^{51}$ ergs,][]{Brown95}. The latter must be revised downward to 
$5 \times 10^{50}$ ergs for an expansion velocity of 15 \kmpers. If this is confirmed, the thermal energy amounts to 20\% to 50\% of the current kinetic energy in the expanding gas.

\citet{Pon16} have modelled the expansion of the overall superbubble assuming an ellipsoidal geometry. Expansion in an exponentially stratified ISM can match the \halpha data for a variety of inclinations onto the Galactic plane \citep{Pon16}, but all models require an internal (uniform) pressure of about $10^4$~cm$^{-3}$~K and an internal temperature of 3 to 4 MK. We find slightly higher values in the X-ray emitting hot gas across the whole Eridanus end of the superbubble, with pressures in the range of $(3.2-7.8) \times 10^4$~cm$^{-3}$~K. The pressure at the closest end of the superbubble towards the \southloop exceeds that in the Local Bubble, which is filled with 1 MK gas with a pressure of $10^4$~cm$^{-3}$~K \citep{Puspitarini14}. We formally find pressures that are at least  three times higher in the \southloop, but we recall that the measurements in the Local Bubble and in \southloop have large uncertainties. An overpressure is consistent with the expansion of the Eridanus tip of the superbubble moving towards us; it will push against the Local Bubble if they are in contact.

The gas in both phases exhibits ten times higher volume densities than were found in the simulations of \citet{Kim17} mentioned above.
The 80~pc size of the X-ray emitting regions implies gas column densities of $(0.6-1.8) \times 10^{18}$ cm$^{-2}$ in the hotter plasma and $(0.4-2.4) \times 10^{19}$ cm$^{-2}$ in the colder plasma. These values compare well with the superbubble simulations of \citet{Krause14} during the short stage when a recent supernova shock approaches the outer shell and drives a peak in X-ray luminosity. At other times, that is, during the megayear-long period that separates supernovae in Eridanus \citep{Voss10}, the simulated column densities are ten times lower than what we measure in Eridanus. However, \citet{Krause14} noted that their simulated X-ray luminosity fades too rapidly after a supernova explosion compared to various observations. A longer delay after the last supernova might therefore still be consistent with the Eridanus X-ray data. Half a million years after an off-centred supernova, the simulations show that the hot gas globally oscillates across the superbubble. The onset of such oscillations could explain the need for a multi-phase hot plasma towards all the sampled areas as well as the presence of hot gas in the \southloop despite the lack of young massive stars. It is so far unclear whether the three hot regions of the \southloop, the enclosure between arcs A and B, and the rest of EXE1 above the \eastshell represent distinct bubbles that are in the process of merging, or whether they represent the dynamical response of the overall superbubble plasma to the sequence of stellar winds and supernovae that have occurred along the stream of blue stars that extends to Orion \citep{Pellizza05,Bouy15}. A joint modelling of the \halpha and X-ray emissions is required to follow the gas cooling inside the superbubble in order to distinguish the origin of the hardness variations across the different zones.

\begin{figure}[p]
\resizebox{0.94\hsize}{!}{\includegraphics{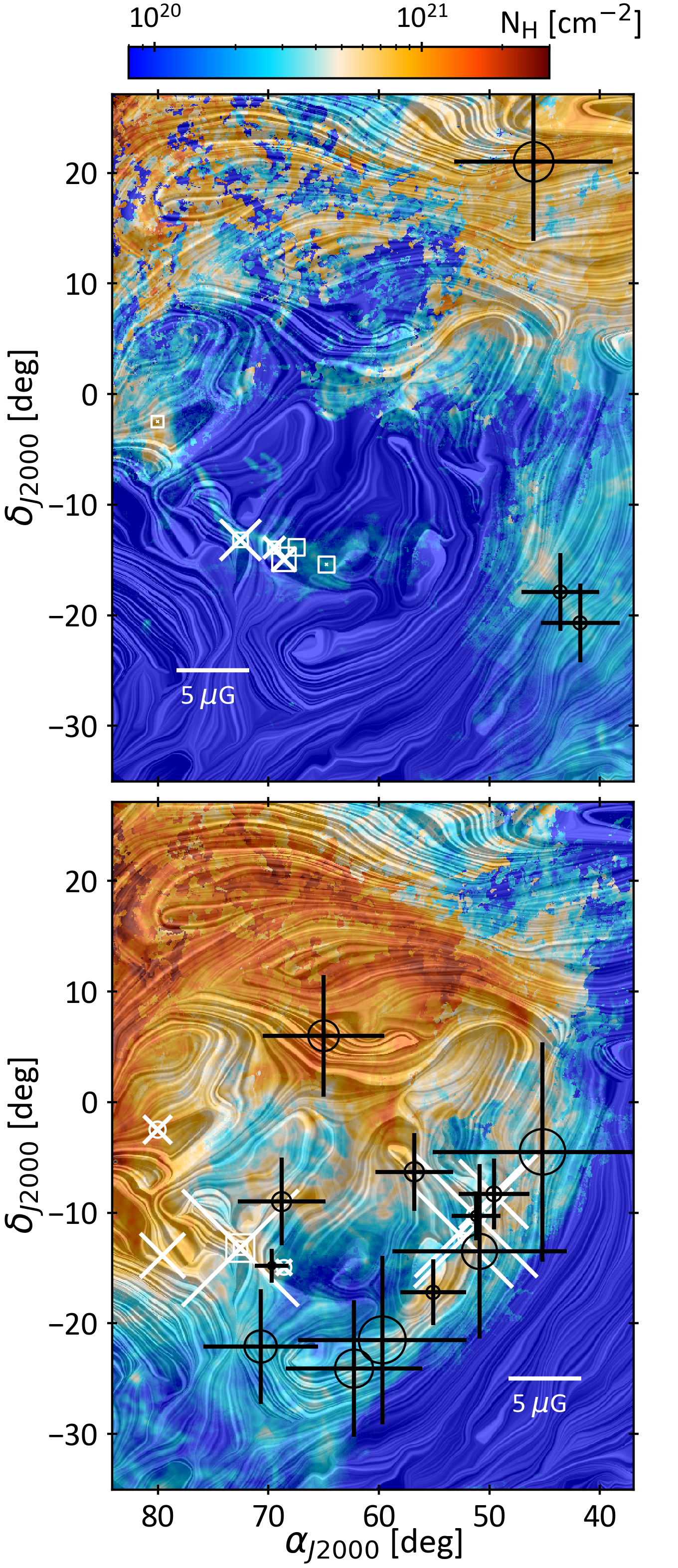}} %0.6 for referee, 0.95 normal
\caption{\hi column densities for mostly negative (\textit{top}) and positive (\textit{bottom}) velocity clouds. The top plot highlights the \eastshell, Eridu cirrus, Cetus and North Taurus clouds, and the \northrim, \westrim, \southloop, and MBM 20 are shown in the bottom plot. The overlaid drapery pattern was produced using the line integral convolution technique \citep[LIC;][]{Cabral93} to show the plane-of-sky magnetic field orientation, \bsky, inferred from the \Planck 353\,GHz polarisation observations. The white crosses show the line-of-sight \blos Zeeman measurements by \citet{Heiles89}, with size proportional to the strength and with 1$\sigma$ error squares or circles for positive or negative values. The black plus signs show the present \bsky estimates based on the angular dispersion in the polarisation data and on the velocity dispersion in the corresponding \hi cloud. The marker size is proportional to the value, and the circle gives the 1$\sigma$ error.}
\label{fig:BMap}
\end{figure}

\section{Magnetic field in the superbubble}
\label{sec:Bfield}

\begin{table*}
\caption{Estimates of the magnetic field strength.}
\centering 
\begin{tabular}{l r@{, }l c r@{ $\pm$ }l r@{ $\pm$ }l r@{ $\pm$ }l r@{ $\pm$ }l r@{ $\pm$ }l r@{ $\pm$ }l} 
\label{tab:BDCF} 
Cloud name & \multicolumn{2}{c}{($\alpha$, $\delta$)} & $\delta \rm{V_c}$ [km s$^{-1}$] & \multicolumn{2}{c}{n$\rm{_H}$ [cm$^{-3}$]} & \multicolumn{2}{c}{$\varsigma_{\psi}$ [deg]} & \multicolumn{2}{c}{$s$ [deg]} & \multicolumn{2}{c}{\bsky [$\mu$G]} & \multicolumn{2}{c}{\blos$^{{a}}$ [$\mu$G]} \\ 
\hline \hline   
\eastshell & (80.0&-2.5) & \multicolumn{9}{}{}{}{}{}{}{}{}{} & 0.1&0.8 \\ 
\eastshell & (67.4&-13.9) & \multicolumn{9}{}{}{}{}{}{}{}{}{} & 0.0&1.1 \\ 
\eastshell & (64.7&-15.4) & \multicolumn{9}{}{}{}{}{}{}{}{}{} & 0.1&1.1 \\ 
\eastshell & (69.5&-13.9) & \multicolumn{9}{}{}{}{}{}{}{}{}{} & 2.1&0.8 \\ 
\eastshell & (72.5&-13.2) & \multicolumn{9}{}{}{}{}{}{}{}{}{} & 3.9&1.0 \\ 
\eastshell & (68.6&-14.9) & \multicolumn{9}{}{}{}{}{}{}{}{}{} & 2.3&1.6 \\ 
\northrim & (56.8&-6.3) & 0.77 & 10&5 & 9.7&0.1 & 14.5&0.8 & 5.3&1.3 \\ 
\northrim & (65.0&6.0) & 0.78 & 22&11 & 9.4&0.1 & 13.1&0.7 & 8.3&2.1 \\ 
\northrim & (80.0&-2.5) &  \multicolumn{9}{}{}{}{}{}{}{}{}{} & -2.7&1.1 \\ 
\northrim & (72.5&-13.2) &  \multicolumn{9}{}{}{}{}{}{}{}{}{} & 11.2&1.9 \\ 
\northrim & (72.5&-13.2) &  \multicolumn{9}{}{}{}{}{}{}{}{}{} & 2.6&0.9 \\ 
\westrim$\!^{b}$ & (49.6&-8.3) & 0.69 & 3&1 & 5.5&0.1 & 8.0&0.8 & 4.8&1.0 & -6.6&0.6 \\ 
\westrim$\!^{b}$ & (51.2&-10.3) & 0.60 & 3&1 & 6.6&0.1 & 9.7&0.8 & 3.3&0.7 & -11.8&1.3 \\ 
\westrim$\!^{b,\ c}$ & (52.3&-12.3) & 1.15 & 3&1 & 6.4&0.1 & 9.4&0.8 & 6.1&1.2 & -9.5&0.7 \\ 
\westrim & (55.1&-17.2) & 0.49 & 5&2 & 5.3&0.1 & 8.0&0.8 & 4.5&0.9 \\ 
\westrim & (50.9&-13.5) & 0.82 & 4&2 & 2.9&0.1 & 4.1&0.8 & 11.9&2.4 \\ 
\westrim & (45.2&-4.5) & 0.80 & 4&2 & 2.3&0.1 & 3.4&0.7 & 15.0&3.1 \\ 
\southloop & (62.2&-24.1) & 0.92 & 3&2 & 3.9&0.1 & 5.7&0.8 & 9.3&2.6 \\ 
\southloop & (59.7&-21.5) & 1.04 & 3&2 & 3.4&0.1 & 5.0&0.8 & 11.5&3.2 \\ 
\southloop & (70.7&-22.1) & 0.87 & 3&2 & 4.0&0.1 & 6.3&0.8 & 7.8&2.2 \\ 
\southloop & (79.6&-13.9) &  \multicolumn{9}{}{}{}{}{}{}{}{}{} & 4.3&0.3 \\ 
G203-37 & (68.8&-9.0) & 0.68 & 30&13 & 13.2&0.1 & 18.1&0.9 & 6.0&1.3 \\ 
MBM 20 & (69.7&-14.8) & 0.25 & 38&15 & 13.9&0.1 & 18.8&0.9 & 2.3&0.5 \\ 
MBM 20 & (68.6&-14.9) &  \multicolumn{9}{}{}{}{}{}{}{}{}{} & -1.5&1.0 \\ 
Eridu & (41.8&-20.7) & 0.54 & 7&2 & 5.8&0.1 & 8.8&0.8 & 5.4&0.9 \\ 
Eridu & (43.6&-17.9) & 0.70 & 7&2 & 7.3&0.1 & 10.7&0.8 & 5.3&0.9 \\ 
North Taurus & (46.0&21.0) & 0.80 & 10&5 & 5.1&0.1 & 7.1&0.8 & 10.8&2.7 \\ 
\end{tabular}
\begin{tablenotes}
\small
\item $^{a}$ From \citet{Heiles89}.
\item $^{b}$ Directions previously studied by \citetalias{Soler18}.
\item $^{c}$ Direction with noisy polarisation data, see text.
\end{tablenotes}
\end{table*}

The magnetic field in the Orion-Eridanus superbubble has recently been analysed by \citet[ \citetalias{Soler18} for short,]{Soler18} using \Planck 2015 polarisation observations at 353\,GHz. Their study highlighted the interaction between the superbubble and the magnetic field, as revealed by the strong polarisation fraction and the low dispersion in polarisation angle along the outer shell. Figure \ref{fig:BMap} shows the orientation of \bsky, the plane-of-sky magnetic field that overlies the \hi column density for negative-velocity structures with $-30 < v_{LSR} < -4$~km/s and for positive-velocity structures with $-4 < v_{LSR} < 25$~km/s. The bottom panel shows that the magnetic field, frozen in the gas, has been swept up by the superbubble expansion. The orientation of \bsky indeed follows the shape of the shell along the \westrim and \northrim. A smaller magnetic loop corresponds to the \southloop. Inside the superbubble, the magnetic field appears to be more disordered, as expected from the activity of stellar winds and past supernovae. Outside the superbubble, at higher latitudes below the Galactic plane (e.g. in the lower right corner of Fig. \ref{fig:BMap}), the projected field lines appear to be smoothly ordered and nearly parallel to the Galactic plane. The expansion of the shock front along the \westrim and part of the \southloop is therefore in a favourable configuration to efficiently compress the external magnetic field.

    \subsection{\bsky estimation method}
\citetalias{Soler18} probed the magnetic field strength in three directions along the \westrim\ by combining \bsky estimates they derived from polarisation data and the Davis-Chandrasekhar-Fermi method \citep[hereafter the DCF method,][]{Davis51, Chandrasekhar53} together with estimates of the \blos field strength along the line of sight obtained by Zeeman splitting \citep{Heiles89}. The three directions are visible in the bottom panel of Fig. \ref{fig:BMap}, together with other Zeeman measurements from \citet{Heiles89}. They found \bsky values of 25 to 87~$\mu$G that far exceed both \blos and the average field strength found in cold atomic clouds in the solar neighbourhood \citep[${\sim} 6\ \mu$G,][]{Heiles05}. The DCF method, however, tends to overestimate the field strength because of line-of-sight and beam averaging. A correction factor of about $\sim$0.5 was advised by \citet{Ostriker01}. It was not used in \citetalias{Soler18} because the probed regions had different physical conditions than in the MHD simulations that were employed to estimate the correction. Instead of applying this factor, we used the modified DCF method described by \cite{Cho16} to probe \bsky in the superbubble. Their method takes into account the reduction in angular dispersion of the magnetic field orientation that is due to averaging effects, which in turn arise from the pile-up of independent eddies along the line of sight. Assuming that the dispersion in \bsky orientation is small and is due to isotropic Alfv\'enic turbulence, their expression to derive the plane-of-sky magnetic field strength is
\begin{equation*}
B_{\rm sky} = \xi \sqrt{4 \pi \rho} \frac{\delta \rm{V}_{c}}{\varsigma_{\psi}}
\end{equation*}
where $\rho$ is the gas mass density, $\delta \rm{V_{c}}$ is the standard deviation of \hi emission line velocities, and $\varsigma_{\psi}$ is the angular dispersion of the local magnetic-field orientations. $\xi$ is a correction factor derived from simulations with values between $\sim$0.7 and $\sim$1. The results presented here are calculated for $\xi=0.7$ because the DCF method tends to overestimate field strengths.

We applied this method towards several directions in the superbubble. They were chosen to exhibit a single dominant velocity structure in the probed area for a reliable evaluation of $\delta \rm{V_{c}}$. This criterion rejected \bsky estimates in broad regions of the \northrim or towards Cetus where two velocity components gather along the lines of sight. We still found two directions that intercept the approaching front of the \northrim without much contamination from other structures. For all directions in our sample, we verified that more than 70\% of the dust thermal emission arises from the cloud of interest to ensure that the measured dispersion in magnetic field orientations strongly relates to that cloud and to limit the overestimation of the field strength because the Planck data are averaged over the entire line of sight. Because of this selection and because of velocity information of the \hi lines, we can associate each magnetic field estimate with a single gas complex, as presented in Sect. \ref{sec:HICO}. The probed directions were also chosen to have a signal-to-noise ratio higher than 3 in polarised intensity, $P/\sigma_P \ge 3$, to derive the polarisation angle dispersion $\varsigma_{\psi}$. This second criterion rejected estimates of \bsky in the \eastshell and in one of the directions analysed by \citetalias{Soler18} towards $(\alpha,\delta)=(52.3,-13.3)$, which they considered at a lower angular resolution to maintain $P/\sigma_P \ge 3$. The resulting \bsky value is displayed in Table \ref{tab:BDCF} but is not used in the following analysis. We note that the measurements of \citetalias{Soler18} correspond to regions where 25\%\ to 45 \% of the dust emission comes from the \northrim and not from the \westrim that they wished to probe. We therefore added directions towards denser parts of the \westrim. 

The \hi line velocities come from the decomposition analysis presented in Sect. \ref{sec:HICO}: they are the central velocities of individual pseudo-Voigt lines fitted against the \hi spectra. \citetalias{Soler18} estimated \bsky in circles with 3\degr\  diameters. In order to avoid velocity gradients or crowding across the test regions, we reduced the diameters to 2\degr. We selected the \hi lines pertaining to the cloud complex to be probed, with peak brightness temperatures above 7~K in order to avoid noisy and poorly detected lines. We computed the standard deviation of the resulting distribution, $\delta \rm{V_{c}}$, weighted by the line brightness temperatures. The results are displayed in Table \ref{tab:BDCF}.

We used two methods to derive the angular dispersion of the local magnetic field orientations, $\varsigma_{\psi}$. The first directly uses the Stokes parameters as defined in \citet{PlanckXXXV16}. The second relies on the structure function of the polarisation angles. Both methods yield consistent results (see Fig. \ref{fig:Bcomp}), and they are detailed in appendix \ref{ap:polar}.

We derived the average mass volume density $\rho$ from the \hi column density assuming a line-of-sight depth. The \southloop forms a nearly complete ring with a thickness of $\sim 10$~pc and a radius of 36~pc at a distance of 200~pc. The \westrim cloud lies along a roughly spherical cap of comparable thickness and with a radius of 90~pc at a distance of 200~pc. Upper limits to the line-of-sight depth across these shells are given by the crossing lengths that graze the inner radius. They correspond to 60 and 100~pc for the \southloop and the \westrim, respectively, but they yield volume densities of $\sim 1-2$~cm$^{-3}$ and \bsky strengths of $\sim 2-4~\mu$G, implying unrealistic Alfv\'en velocities that exceed the expansion velocity of the superbubble. When we take the thickness of the cap as a lower limit to the depth, we obtain average volume densities of $\sim 20$~cm$^{-3}$ , which is rather high for such diffuse atomic clouds. The low \hi emission brightness temperatures do not support large amounts of cold optically thick gas along the lines of sight. We therefore assumed mean depths of $40 \pm 20$~pc and $50 \pm 20$~pc for the \southloop and the \westrim. For the other clouds, we assumed that the transverse size applies in the third dimension, as suggested by the compactness of MBM 20 and G203-37 and the filamentary structure of Eridu.  We assumed a mean weight $\mu=1.36 m_{\rm H}$ per hydrogen atom in the gas. The resulting gas number densities, $n_{\rm H}$, are listed in Table \ref{tab:BDCF}. The values range between 3 and 10 cm$^{-3}$ in the shells, in agreement with superbubble simulations \citep{Kim17}. Twice higher values are found in the \northrim, as expected from the optically thicker conditions in \hi and the dark neutral gas (see Paper II). The highest values correspond to the two compact molecular clouds, MBM 20 and G203-37.

    \subsection{\bsky estimates}
    
The derived values of \bsky are listed in Table \ref{tab:BDCF} and displayed in Figure \ref{fig:BMap}. Our assumption of a low dispersion in polarisation orientation angle is valid: it is below the 25$\degr$ limit proposed by \citet{Ostriker01} from simulations, so that the ordered component of the magnetic field is much larger than the turbulent component. We obtain \bsky strengths ranging from $\sim3$ to $\sim15~\mu$G towards the superbubble clouds. We can compare these values with the \citet{Sofue19} results, which are based on synchrotron emission and Faraday rotation measures. They have derived all-sky maps of \blos and $B_{\rm tot} = (B_{\rm sky}^2+B_{\rm los}^2)^{1/2}$, assuming equipartition between the magnetic and cosmic-ray energy densities. In the total magnetic field map, the Orion-Eridanus superbubble stands out with field strengths of 10 to 14~$\mu$G compared to a local average of $7~\mu$G. For the two directions in our sample where \btot is available, we obtain values of 8.2$\pm$1.2 and 12.3$\pm 1.5~\mu$G, in good agreement with the $12~\mu$G value found at these locations in the \citet{Sofue19} map. We further discuss magnetic compression along the outer shell of the superbubble in the next section.
We find \bsky strengths that are a factor of 17 to 18 times lower than the \citetalias{Soler18} estimates in the same directions along the \westrim. The difference stems from both higher values of the line-of-sight depth and the \citet{Cho16} correction. The effect of the modified DCF method alone would lower their magnetic strengths by a factor of 5 in these directions. 

We have an estimate of the magnetic field in, or close to the Local Bubble wall, towards MBM 20. The \blos value from \citet{Heiles89} is consistent with that of \citet{Xu19}. Our estimate of \bsky of $2.3\pm0.5~\mu$G is lower than the  $=8^{+5}_{-3}~\mu$G derived by \citet{Andersson06} using the DCF method, but this difference may originate from the fact that we have probed the magnetic field in the atomic gas phase, whereas they performed their measurement in the molecular gas, which generally exhibits stronger fields \citep{Crutcher2012}.

\begin{figure}
\centering
\resizebox{\hsize}{!}{\includegraphics{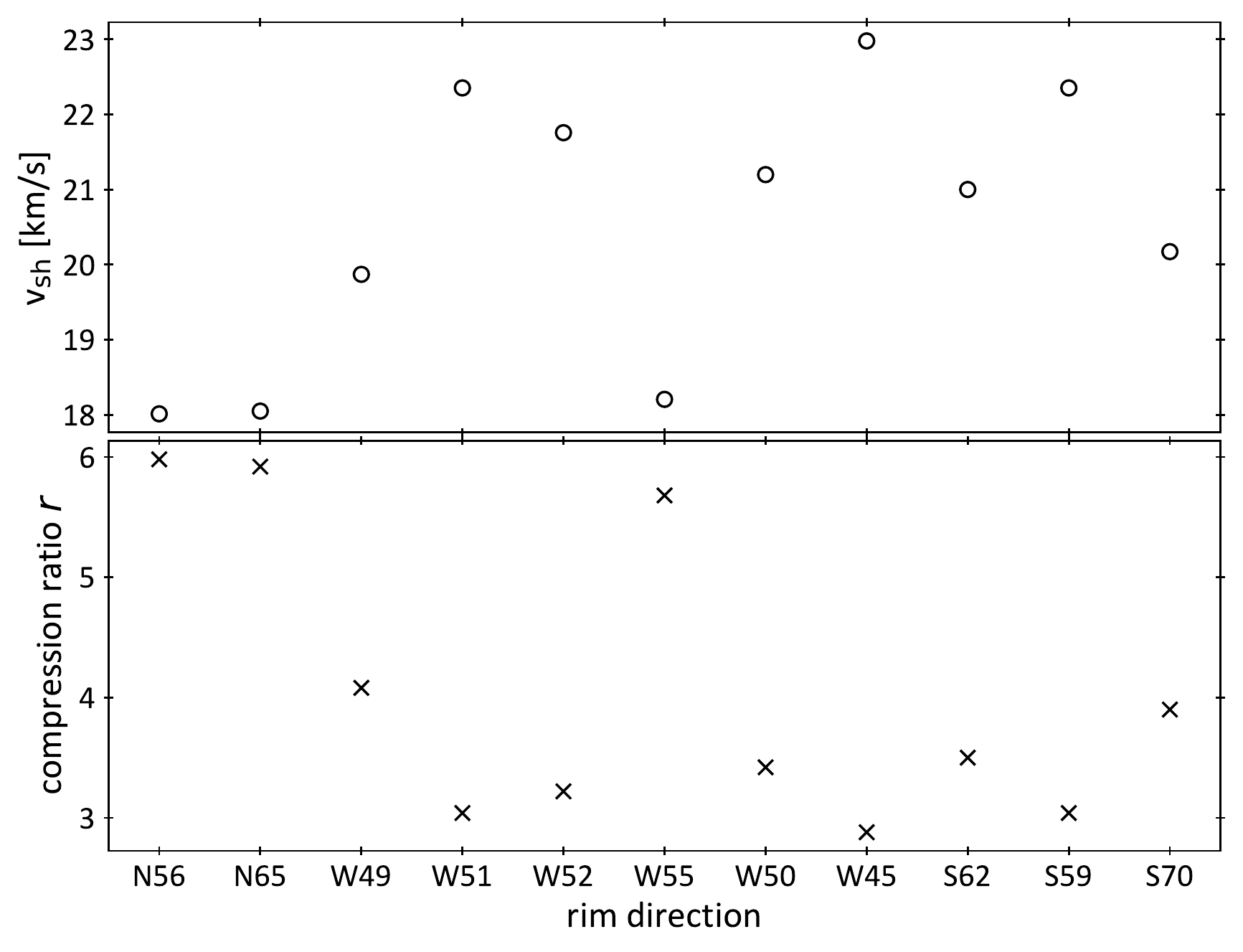}}
\caption{Estimates of the velocity and compression ratio of an 8000~K isothermal shock constrained by the downstream gas densities and magnetic field strengths listed in Table \ref{tab:BDCF} and by the LSR expansion velocity of 15~km/s of the downstream \hi shells. The different directions are noted with their truncated right ascension, and the letters N, W, or S for the \northrim, \westrim, and \southloop rims, respectively.}
\label{fig:choc}
\end{figure}

    \subsection{Outer shock velocity and compression ratio}
    \label{sub:choc}
We find that the fields along the outer shell of the superbubble (along the \westrim and the \southloop) span \bsky values from 5 to about 15~$\mu$G, which means that the total field downstream of the expanding shock wave exceeds the mean total field $B_{tot}=7~\mu$G that prevails outside the superbubble, at high Galactic latitudes \citep{Sofue19}. \citet{Xu19} reported a similar enhancement by a factor of 2 in the shell of the Gum nebula, using rotation measures from pulsars. The \bsky strengths towards the two \northrim directions mostly probe the magnetic field in the front part of the outer shell. Their values of $5.3 \pm 1.3~\mu$G and $8.3 \pm 2.1~\mu$G are much higher than the $\sim 1~\mu$G field strengths found in \blos in the same directions \citep{Sofue19}, therefore the field lines are highly inclined to the lines of sight and to the shock velocity. A similar configuration has been observed in the wall of the Local Bubble, where \citet{Xu19} have found a value of \blos $=0.5-2~\mu$G much lower than the \bsky $=8^{+5}_{-3}~\mu$G derived by \citet{Andersson06} using the DCF method.

These low strength \blos ($\lesssim 5~\mu$G) are also consistent with estimates from rotation measures of pulsars. The Australia telescope national facility (ATNF) pulsar catalogue lists 11 pulsars in the region of analysis, 2 of which are in the distance range of the Orion-Eridanus superbubble : J0452-1759 at 400~pc towards ($\alpha, \delta$)=(73.1\degr,-18\degr) and J0459-0210 at 160~pc towards ($\alpha, \delta$)=(75\degr,-2.2\degr). The ratio of the rotation measure to the dispersion measure gives the line-of-sight integrated \blos, weighted by the thermal electron density along the line of sight. The \blos strengths of $0.34 \pm 0.01$ and 1.05$\pm$0.53~$\mu$G towards the two pulsars are consistent with the Zeeman measurements in the atomic phase, but the latter mostly samples the magnetic field in the Local Bubble because of the close distance to the J0459-0210 pulsar.

As indicated by the dust polarisation orientation that closely delineates the outer shell in Fig. \ref{fig:BMap} and by the large inclination of the field lines to the shock velocity in the front wall, the measured field strengths likely result from magnetic compression of the external field by the expanding shock wave. The atomic gas shells also result from compression and rapid cooling of the swept-up gas. For a shock velocity of $\sim$20~km/s, a compression ratio of 4, a mean gas density of 5~cm$^{-3}$, and the local interstellar cooling function \citep{Richings14}, the radiative cooling length scale downstream of the shock is smaller than 0.1 pc. We can therefore consider the shock as isothermal to infer its compression ratio, $r$, and its velocity, $v_{sh}$, from the downstream data. 

The expansion velocity, $v_{exp}$, of the downstream \hi shells in the local standard of rest is related to the shock velocity as $v_{exp} = v_{sh}(1 - \frac{1}{r})$. Mass conservation and the isothermal condition relate the mass densities, $\rho$, gas velocities in the shock frame, $u$, and gas pressures, $p$, on both sides of the shock as $r = \rho_d/\rho_u = p_d/p_u= u_u/u_d $ , where the subscripts $u$ and $d$  note the upstream and downstream media, respectively. Magnetic flux conservation  relates the magnetic field strengths perpendicular to the shock velocity as $B_{d{\perp}}/B_{u{\perp}} = r$. Using these jump relations to eliminate the upstream variables that are not measured, we can write the momentum conservation equation as
\begin{equation}
\label{eq:choc}
    \frac{p_d}{r} + \frac{r}{(r-1)^2}\rho_d v_{exp}^2 + \frac{B_{d\perp}^2}{2 \mu_0 r^2} = p_d +  \frac{1}{(r-1)^2}\rho_d v_{exp}^2 + \frac{B_{d\perp}^2}{2 \mu_0}
.\end{equation}
We assumed a temperature $T_u=T_d=8000$~K typical of the warm atomic gas to relate the downstream pressure $p_d$ to the mean downstream gas densities listed in Table \ref{tab:BDCF}. We numerically solved equation \ref{eq:choc} for the compression ratio $r$ in each direction.

The magnetic field component perpendicular to the shock velocity is close to \bsky towards the two \northrim directions that probe the front wall, whereas both \bsky and \blos contribute to $B_{d{\perp}}$ along the western and southern rims. We took the corresponding \bsky and \blos values from Table \ref{tab:BDCF} for the \northrim, \westrim, and \southloop data points. We completed the \blos dataset with RM and synchrotron estimates from \citet{Sofue19}, who found $0\lesssim $\blos$\lesssim 3\,\mu$G along the \westrim and $3\lesssim $\blos$\lesssim 5\,\mu$G along the \southloop. We note that the resulting shock velocities and compression ratios vary by less than 5\% and 10\%, respectively, when we include or exclude the \citet{Sofue19} \blos values in the calculations. The results obtained for $v_{exp}= 15$~km/s are shown in Fig. \ref{fig:choc}.

The data point with little dispersion to an outer shock velocity close to 20~km/s and a compression ratio near 4. The downstream Alfv\'en velocities of 3 to 14~km/s are compatible with magnetic compression of external fields with $B{u{\perp}}$ strengths ranging from 1 to 5~$\mu$G around the superbubble. The data require upstream pressures of 6000 to 11000~K cm$^{-3}$ along the western and southern rims that are compatible with the Local Bubble environment. Higher upstream pressures of 13000 and 30000~K are required to explain the higher gas densities seen towards the two \northrim directions, but the downstream densities in these directions have larger uncertainties because the thickness of the front shell is unknown. 

\section{Conclusions}
\label{sec:concl}

\subsection{Cloud shells and outer shock expansion}
We have studied the distribution in position and velocity of \hi and CO emission lines towards the Eridanus part of the Orion-Eridanus superbubble. We decomposed the gas distribution into individual cloud complexes that appear as rings or shells of neutral gas that flank the \halpha recombination arcs. Arc B extends alongside the \westrim, Arc C along a part of the \southloop, Arc A is related to the molecular part of the \northrim, and the \eastshell is parallel to an unnamed \halpha arc that is visible at negative velocities. Visual extinctions of 1-2 magnitudes along the northwest part of the outer rim hinder the detection of the \halpha recombination arcs. 

The clouds also appear as separate entities in distance in 3D dust reddening maps. Their location, $\sim$150 pc to $\sim$250 pc away from the Sun, confirms that the superbubble is oriented with its far end towards Orion and its close end towards Eridanus. The data suggest that the \southloop is the closest and approaching end of the superbubble. The central velocities of the \hi lines trace the bulk motions of the gas. Their distribution highlights another coherent ring at about +9 \kmpers that likely marks the outer shell of the superbubble. We find expansion velocities of 10-15 \kmpers about this rest velocity. These motions are slower than the 40 \kmpers proposed by \citet{Brown95} from the highest velocities recorded in the wings of the \hi lines, but they are consistent with the values of 15-23 \kmpers found in \halpha \citep{Reynolds1979}. 

The gas shells likely result from compression and rapid cooling of the gas swept up by the outer shock wave of the superbubble. The average gas densities, the short radiative cooling length scale, and the enhanced strength and orientation of the magnetic fields along the shells support this interpretation. The measured expansion velocity of 15~\kmpers therefore corresponds to the downstream gas velocity, which differs from the shock velocity itself. We have inferred a shock velocity of 18-23 \kmpers for an isothermal shock constrained by the downstream data on gas and magnetic fields and for a gas temperature of 8000~K. The compression ratio of 3 to 6 implies magnetic and gas pressures upstream of the shock that are consistent with the Local Bubble values.

\subsection{Hot gas interior}
We have used the ROSAT data in the three energy bands around 0.25, 0.75, and 1.5 keV and calculated the optical depth in these bands for the individual gas shells and for the total gas column densities measured along the lines of sight. We used this information to study the hot gas inside the superbubble and to locate the clouds relative to the hot gas. We also modelled the X-ray spectra in ten specific zones using mekal emission models and the gas absorption to infer the thermal properties of the hot gas.

We find that the \eastshell nicely bounds the emission at 0.75 and 1.5 keV. This boundary is not entirely due to absorption because the cloud is optically thin to 1.5~keV X-rays and partially thin ($\tau_X$ of 0.1-0.4) at 0.75~keV. The cloud is optically thick at 0.25~keV, but it does not heavily absorb the softest X-rays, meaning that it lies inside the superbubble rather than in front of the hot gas. In contrast, the \southloop is closer and absorbs the 0.25~keV emission. We used our gas data to revisit the origin of the soft X-ray emission towards the \southloop. As suggested by \citet{Snowden95}, we find that significant 0.25~keV emission remains inside the loop after the foreground emission from the Local Bubble and the background intensity are subtracted. The background intensity is the intensity that is expected from the Galactic halo and is modulated by the total gas absorption. 

Taking the \southloop absorption into account, we find no clear spectral difference between the EXE1 region north of the \eastshell and the EXE2 region towards the \southloop. The spectral dichotomy that is visible on the maps is largely induced by the \southloop absorption. A single-temperature plasma cannot match the X-ray spectra across the three bands in any of the sampled directions, however. Two hot phases are required, with respective temperatures of 3-9~MK and 0.3-1.2~MK under the assumption of a uniform pressure across both phases. The derived pressure of $(3-8) \times 10^4$~cm$^{-3}$~K exceeds the pressure of the Local Bubble of $10^4$~cm$^{-3}$~K \citep{Puspitarini14}. The need for at least two plasma phases might indicate global plasma oscillations across the superbubble that are caused by its complex interior history \citep{Krause14}. In addition, the significant emission hardening seen towards the region that is enclosed between the \halpha arcs A and B corresponds to a higher pressure and hotter temperature that support the hypothesis of a younger age and faster expansion of this sub-region \citep{Heiles99}. 

The origin of the \southloop ring is unclear. A wind-blown bubble seems unlikely because of the lack of massive stars in this direction \citep{Burrows93}. We find no gradient in temperature or pressure between the \southloop and adjacent regions that would support a cooling flow of hot gas that leaks at the rear of the superbubble \citep{Heiles99}. The gas ring, filled with plasma that is a few million degrees hot, is consistent with an old supernova remnant. The progenitor star may have belonged to the Cassiopeia-Taurus OB association, which extends from 130 to 300~pc \citep{DeZeeuw99,Pellizza05} in these directions. 
The more sensitive and refined X-ray maps that will soon be obtained by the extended Roentgen survey with an imaging telescope array (e-ROSITA) will spatially and spectrally constrain the heterogeneity of the hot gas phases that fill the superbubble better. This will enable us to reconstruct the possible sequence of supernovae that have blown this structure.

\subsection{Magnetic field in the superbubble}
We have used \Planck polarised dust emission and the gas dynamics information from the \hi line decomposition to study the magnetic field in the superbubble. We probed the plane-of-sky magnetic field, \bsky, using the modified Davis-Chandrasekhar-Fermi method (DCF method) proposed by \citet{Cho16}. We have obtained magnetic strengths 17 times lower than previous estimates along the \westrim \citep{Soler18}. The difference stems from (i) the selection of regions with little \hi confusion along the lines of sight, (ii) refined estimates of the gas shell depths along the lines of sight, and (iii) the correction of the dispersion in polarisation angles due to the pile-up of independent eddies along the line of sight.

The expanding superbubble has likely swept up and compressed the external magnetic field component perpendicular to the shock velocity. The polarisation alignment with the gas shell along the \westrim and \southloop, the enhanced \bsky strengths that reach up to 15 $\mu$G, and the estimates of the Alfv\'en velocities in the downstream gas support this interpretation. The high ratio found between the \bsky and \blos strengths towards approaching gas structures in the \northrim is also consistent with magnetic compression in the front wall of the superbubble. The DCF method only provides approximate field strengths, however. The $0.7 \lesssim \xi \lesssim 1$ span in correction factor and the uncertainty on the line-of-sight depths lead to large uncertainties in the resulting \bsky. Furthermore, large fractions of the superbubble could not be probed in $B$ because of the limited signal-to-noise ratio in polarisation. Future dust polarisation observations at higher sensitivity and angular resolution will enable a finer sampling of \bsky and will allow additional corrections, for instance of the polarisation smoothing in the telescope beam and the effect of small-scale ordered density gradients across the aperture used to measure the magnetic dispersion \citep{Hildebrand09, Houde09}.

\begin{acknowledgements}
We acknowledge the financial support from the ANR-DFG grant CRiBs (ANR-15-CE31-0019-01) for this work. We also made use of XSPEC \footnote{See \url{ https://heasarc.gsfc.nasa.gov/xanadu/xspec/manual/XspecManual.html}}. We acknowledge the \Planck collaboration for the publicly available data through the Planck Legacy Archive. JDS acknowledges funding from the European Research Council under the Horizon 2020 Framework Program via the ERC Consolidator Grant CSF-648505. We thank A. Bracco for the useful comments and discussions. We also thank the referee for the valuable comments and suggestions.
\end{acknowledgements}

\bibliographystyle{aa}
\bibliography{References}

\begin{thebibliography}{58}
\expandafter\ifx\csname natexlab\endcsname\relax\def\natexlab#1{#1}\fi

\bibitem[{{Andersson} \& {Potter}(2006)}]{Andersson06}
{Andersson}, B.~G. \& {Potter}, S.~B. 2006, \apj, 640, L51

\bibitem[{{Arnaud}(1996)}]{Arnaud1996}
{Arnaud}, K.~A. 1996, in Astronomical Society of the Pacific Conference Series,
  Vol. 101, Astronomical Data Analysis Software and Systems V, ed. G.~H.
  {Jacoby} \& J.~{Barnes}, 17

\bibitem[{{Bally}(2008)}]{Bally2008}
{Bally}, J. 2008, {Overview of the Orion Complex}, Vol.~4 (ASP Monograph
  Publications), 459

\bibitem[{{Bouy} \& {Alves}(2015)}]{Bouy15}
{Bouy}, H. \& {Alves}, J. 2015, \aap, 584, A26

\bibitem[{{Brown} {et~al.}(1995){Brown}, {Hartmann}, \& {Burton}}]{Brown95}
{Brown}, A.~G.~A., {Hartmann}, D., \& {Burton}, W.~B. 1995, \aap, 300, 903

\bibitem[{{Burrows} {et~al.}(1993){Burrows}, {Singh}, {Nousek}, {Garmire}, \&
  {Good}}]{Burrows93}
{Burrows}, D.~N., {Singh}, K.~P., {Nousek}, J.~A., {Garmire}, G.~P., \& {Good},
  J. 1993, \apj, 406, 97

\bibitem[{{Cabral} \& {Leedom}(1993)}]{Cabral93}
{Cabral}, B. \& {Leedom}, L.~C. 1993, in Special Interest Group on GRAPHics and
  Interactive Techniques Proc., 263

\bibitem[{{Chandrasekhar} \& {Fermi}(1953)}]{Chandrasekhar53}
{Chandrasekhar}, S. \& {Fermi}, E. 1953, \apj, 118, 113

\bibitem[{{Cho} \& {Yoo}(2016)}]{Cho16}
{Cho}, J. \& {Yoo}, H. 2016, \apj, 821, 21

\bibitem[{{Crutcher}(2012)}]{Crutcher2012}
{Crutcher}, R.~M. 2012, \araa, 50, 29

\bibitem[{{Dame} {et~al.}(2001){Dame}, {Hartmann}, \& {Thaddeus}}]{Dame2001}
{Dame}, T.~M., {Hartmann}, D., \& {Thaddeus}, P. 2001, \apj, 547, 792

\bibitem[{{Dame} \& {Thaddeus}(2004)}]{Dame2004}
{Dame}, T.~M. \& {Thaddeus}, P. 2004, in Astronomical Society of the Pacific
  Conference Series, Vol. 317, Milky Way Surveys: The Structure and Evolution
  of our Galaxy, ed. D.~{Clemens}, R.~{Shah}, \& T.~{Brainerd}, 66

\bibitem[{{Davis}(1951)}]{Davis51}
{Davis}, L. 1951, Physical Review, 81, 890

\bibitem[{{de Zeeuw} {et~al.}(1999){de Zeeuw}, {Hoogerwerf}, {de Bruijne},
  {Brown}, \& {Blaauw}}]{DeZeeuw99}
{de Zeeuw}, P.~T., {Hoogerwerf}, R., {de Bruijne}, J.~H.~J., {Brown}, A.~G.~A.,
  \& {Blaauw}, A. 1999, \aj, 117, 354

\bibitem[{{Dolan} \& {Mathieu}(2002)}]{Dolan02}
{Dolan}, C.~J. \& {Mathieu}, R.~D. 2002, \aj, 123, 387

\bibitem[{{Finkbeiner}(2003)}]{Finkbeiner2003}
{Finkbeiner}, D.~P. 2003, \apj S, 146, 407

\bibitem[{{Green} {et~al.}(2018){Green}, {Schlafly}, {Finkbeiner}, {Rix},
  {Martin}, {Burgett}, {Draper}, {Flewelling}, {Hodapp}, {Kaiser}, {Kudritzki},
  {Magnier}, {Metcalfe}, {Tonry}, {Wainscoat}, \& {Waters}}]{Green18}
{Green}, G.~M., {Schlafly}, E.~F., {Finkbeiner}, D., {et~al.} 2018, \mnras,
  478, 651

\bibitem[{{Guo} {et~al.}(1995){Guo}, {Burrows}, {Sanders}, {Snowden}, \&
  {Penprase}}]{Guo95}
{Guo}, Z., {Burrows}, D.~N., {Sanders}, W.~T., {Snowden}, S.~L., \& {Penprase},
  B.~E. 1995, \apj, 453, 256

\bibitem[{{Heiles}(1989)}]{Heiles89}
{Heiles}, C. 1989, \apj, 336, 808

\bibitem[{{Heiles} \& {Crutcher}(2005)}]{Heiles05}
{Heiles}, C. \& {Crutcher}, R. 2005, {Magnetic Fields in Diffuse HI and
  Molecular Clouds}, Vol. 664 (Springer, Berlin, Heidelberg), 137

\bibitem[{{Heiles} {et~al.}(1999){Heiles}, {Haffner}, \& {Reynolds}}]{Heiles99}
{Heiles}, C., {Haffner}, L.~M., \& {Reynolds}, R.~J. 1999, in Astronomical
  Society of the Pacific Conference Series, Vol. 168, New Perspectives on the
  Interstellar Medium, ed. A.~R. {Taylor}, T.~L. {Landecker}, \& G.~{Joncas},
  211

\bibitem[{{HI4PI Collaboration} {et~al.}(2016){HI4PI Collaboration}, {Ben
  Bekhti}, {Fl{\"o}er}, {Keller}, {Kerp}, {Lenz}, {Winkel}, {Bailin},
  {Calabretta}, {Dedes}, {Ford}, {Gibson}, {Haud}, {Janowiecki}, {Kalberla},
  {Lockman}, {McClure-Griffiths}, {Murphy}, {Nakanishi}, {Pisano}, \&
  {Staveley-Smith}}]{HI4PI}
{HI4PI Collaboration}, {Ben Bekhti}, N., {Fl{\"o}er}, L., {et~al.} 2016, \aap,
  594, A116

\bibitem[{{Hildebrand} {et~al.}(2009){Hildebrand}, {Kirby}, {Dotson}, {Houde},
  \& {Vaillancourt}}]{Hildebrand09}
{Hildebrand}, R.~H., {Kirby}, L., {Dotson}, J.~L., {Houde}, M., \&
  {Vaillancourt}, J.~E. 2009, \apj, 696, 567

\bibitem[{{Houde} {et~al.}(2009){Houde}, {Vaillancourt}, {Hildebrand},
  {Chitsazzadeh}, \& {Kirby}}]{Houde09}
{Houde}, M., {Vaillancourt}, J.~E., {Hildebrand}, R.~H., {Chitsazzadeh}, S., \&
  {Kirby}, L. 2009, \apj, 706, 1504

\bibitem[{{Johnson}(1978)}]{Johnson78}
{Johnson}, P.~G. 1978, \mnras, 184, 727

\bibitem[{{Kim} {et~al.}(2017){Kim}, {Ostriker}, \& {Raileanu}}]{Kim17}
{Kim}, C.-G., {Ostriker}, E.~C., \& {Raileanu}, R. 2017, \apj, 834, 25

\bibitem[{{Krause} {et~al.}(2014){Krause}, {Diehl}, {B{\"o}hringer},
  {Freyberg}, \& {Lubos}}]{Krause14}
{Krause}, M., {Diehl}, R., {B{\"o}hringer}, H., {Freyberg}, M., \& {Lubos}, D.
  2014, \aap, 566, A94

\bibitem[{{Krause} {et~al.}(2013){Krause}, {Fierlinger}, {Diehl}, {Burkert},
  {Voss}, \& {Ziegler}}]{Krause13}
{Krause}, M., {Fierlinger}, K., {Diehl}, R., {et~al.} 2013, \aap, 550, A49

\bibitem[{{Lallement} {et~al.}(2018){Lallement}, {Capitanio}, {Ruiz-Dern},
  {Danielski}, {Babusiaux}, {Vergely}, {Elyajouri}, {Arenou}, \&
  {Leclerc}}]{Lallement18}
{Lallement}, R., {Capitanio}, L., {Ruiz-Dern}, L., {et~al.} 2018, \aap, 616,
  A132

\bibitem[{{Liedahl} {et~al.}(1995){Liedahl}, {Osterheld}, \&
  {Goldstein}}]{Liedahl95}
{Liedahl}, D.~A., {Osterheld}, A.~L., \& {Goldstein}, W.~H. 1995, \apj, 438,
  L115

\bibitem[{{Magnani} {et~al.}(1985){Magnani}, {Blitz}, \& {Mundy}}]{Magnani1985}
{Magnani}, L., {Blitz}, L., \& {Mundy}, L. 1985, \apj, 295, 402

\bibitem[{{Mewe} {et~al.}(1985){Mewe}, {Gronenschild}, \& {van den
  Oord}}]{Mewe85}
{Mewe}, R., {Gronenschild}, E.~H.~B.~M., \& {van den Oord}, G.~H.~J. 1985,
  \aaps, 62, 197

\bibitem[{{Nguyen} {et~al.}(2018){Nguyen}, {Dawson}, {Miville-Desch{\^e}nes},
  {Tang}, {Li}, {Heiles}, {Murray}, {Stanimirovi{\'c}}, {Gibson},
  {McClure-Griffiths}, {Troland}, {Bronfman}, \& {Finger}}]{Nguyen18}
{Nguyen}, H., {Dawson}, J.~R., {Miville-Desch{\^e}nes}, M.~A., {et~al.} 2018,
  \apj, 862, 49

\bibitem[{{Ochsendorf} {et~al.}(2015){Ochsendorf}, {Brown}, {Bally}, \&
  {Tielens}}]{Ochsendorf15}
{Ochsendorf}, B.~B., {Brown}, A. G.~A., {Bally}, J., \& {Tielens}, A. G.~G.~M.
  2015, \apj, 808, 111

\bibitem[{{Ostriker} {et~al.}(2001){Ostriker}, {Stone}, \&
  {Gammie}}]{Ostriker01}
{Ostriker}, E.~C., {Stone}, J.~M., \& {Gammie}, C.~F. 2001, \apj, 546, 980

\bibitem[{{Pellizza} {et~al.}(2005){Pellizza}, {Mignani}, {Grenier}, \&
  {Mirabel}}]{Pellizza05}
{Pellizza}, L.~J., {Mignani}, R.~P., {Grenier}, I.~A., \& {Mirabel}, I.~F.
  2005, \aap, 435, 625

\bibitem[{{Planck Collaboration} {et~al.}(2014){Planck Collaboration},
  {Abergel}, {Ade}, {Aghanim}, {Alves}, {Aniano}, {Armitage-Caplan}, {Arnaud},
  {Ashdown}, {Atrio-Barandela}, {Aumont}, {Baccigalupi}, {Banday}, {Barreiro},
  {Bartlett}, {Battaner}, {Benabed}, {Beno{\^\i}t}, {Benoit-L{\'e}vy},
  {Bernard}, {Bersanelli}, {Bielewicz}, {Bobin}, {Bock}, {Bonaldi}, {Bond},
  {Borrill}, {Bouchet}, {Boulanger}, {Bridges}, {Bucher}, {Burigana}, {Butler},
  {Cardoso}, {Catalano}, {Chamballu}, {Chary}, {Chiang}, {Chiang},
  {Christensen}, {Church}, {Clemens}, {Clements}, {Colombi}, {Colombo},
  {Combet}, {Couchot}, {Coulais}, {Crill}, {Curto}, {Cuttaia}, {Danese},
  {Davies}, {Davis}, {de Bernardis}, {de Rosa}, {de Zotti}, {Delabrouille},
  {Delouis}, {D{\'e}sert}, {Dickinson}, {Diego}, {Dole}, {Donzelli},
  {Dor{\'e}}, {Douspis}, {Draine}, {Dupac}, {Efstathiou}, {En{\ss}lin},
  {Eriksen}, {Falgarone}, {Finelli}, {Forni}, {Frailis}, {Fraisse},
  {Franceschi}, {Galeotta}, {Ganga}, {Ghosh}, {Giard}, {Giardino},
  {Giraud-H{\'e}raud}, {Gonz{\'a}lez-Nuevo}, {G{\'o}rski}, {Gratton},
  {Gregorio}, {Grenier}, {Gruppuso}, {Guillet}, {Hansen}, {Hanson}, {Harrison},
  {Helou}, {Henrot-Versill{\'e}}, {Hern{\'a}ndez- Monteagudo}, {Herranz},
  {Hildebrandt}, {Hivon}, {Hobson}, {Holmes}, {Hornstrup}, {Hovest},
  {Huffenberger}, {Jaffe}, {Jaffe}, {Jewell}, {Joncas}, {Jones}, {Juvela},
  {Keih{\"a}nen}, {Keskitalo}, {Kisner}, {Knoche}, {Knox}, {Kunz},
  {Kurki-Suonio}, {Lagache}, {L{\"a}hteenm{\"a}ki}, {Lamarre}, {Lasenby},
  {Laureijs}, {Lawrence}, {Leonardi}, {Le{\'o}n-Tavares}, {Lesgourgues},
  {Levrier}, {Liguori}, {Lilje}, {Linden-V{\o}rnle}, {L{\'o}pez-Caniego},
  {Lubin}, {Mac{\'\i}as-P{\'e}rez}, {Maffei}, {Maino}, {Mandolesi}, {Maris},
  {Marshall}, {Martin}, {Mart{\'\i}nez-Gonz{\'a}lez}, {Masi}, {Massardi},
  {Matarrese}, {Matthai}, {Mazzotta}, {McGehee}, {Melchiorri}, {Mendes},
  {Mennella}, {Migliaccio}, {Mitra}, {Miville-Desch{\^e}nes}, {Moneti},
  {Montier}, {Morgante}, {Mortlock}, {Munshi}, {Murphy}, {Naselsky}, {Nati},
  {Natoli}, {Netterfield}, {N{\o}rgaard-Nielsen}, {Noviello}, {Novikov},
  {Novikov}, {Osborne}, {Oxborrow}, {Paci}, {Pagano}, {Pajot}, {Paladini},
  {Paoletti}, {Pasian}, {Patanchon}, {Perdereau}, {Perotto}, {Perrotta},
  {Piacentini}, {Piat}, {Pierpaoli}, {Pietrobon}, {Plaszczynski},
  {Pointecouteau}, {Polenta}, {Ponthieu}, {Popa}, {Poutanen}, {Pratt},
  {Pr{\'e}zeau}, {Prunet}, {Puget}, {Rachen}, {Reach}, {Rebolo}, {Reinecke},
  {Remazeilles}, {Renault}, {Ricciardi}, {Riller}, {Ristorcelli}, {Rocha},
  {Rosset}, {Roudier}, {Rowan- Robinson}, {Rubi{\~n}o-Mart{\'\i}n}, {Rusholme},
  {Sandri}, {Santos}, {Savini}, {Scott}, {Seiffert}, {Shellard}, {Spencer},
  {Starck}, {Stolyarov}, {Stompor}, {Sudiwala}, {Sunyaev}, {Sureau}, {Sutton},
  {Suur-Uski}, {Sygnet}, {Tauber}, {Tavagnacco}, {Terenzi}, {Toffolatti},
  {Tomasi}, {Tristram}, {Tucci}, {Tuovinen}, {T{\"u}rler}, {Umana},
  {Valenziano}, {Valiviita}, {Van Tent}, {Verstraete}, {Vielva}, {Villa},
  {Vittorio}, {Wade}, {Wandelt}, {Welikala}, {Ysard}, {Yvon}, {Zacchei}, \&
  {Zonca}}]{PlanckXI14}
{Planck Collaboration}, {Abergel}, A., {Ade}, P.~A.~R., {et~al.} 2014, \aap,
  571, A11

\bibitem[{{Planck Collaboration} {et~al.}(2015{\natexlab{a}}){Planck
  Collaboration}, {Ade}, {Aghanim}, {Alina}, {Alves}, {Armitage-Caplan},
  {Arnaud}, {Arzoumanian}, {Ashdown}, {Atrio- Barandela}, {Aumont},
  {Baccigalupi}, {Banday}, {Barreiro}, {Battaner}, {Benabed},
  {Benoit-L{\'e}vy}, {Bernard}, {Bersanelli}, {Bielewicz}, {Bock}, {Bond},
  {Borrill}, {Bouchet}, {Boulanger}, {Bracco}, {Burigana}, {Butler}, {Cardoso},
  {Catalano}, {Chamballu}, {Chary}, {Chiang}, {Christensen}, {Colombi},
  {Colombo}, {Combet}, {Couchot}, {Coulais}, {Crill}, {Curto}, {Cuttaia},
  {Danese}, {Davies}, {Davis}, {de Bernardis}, {de Gouveia Dal Pino}, {de
  Rosa}, {de Zotti}, {Delabrouille}, {D{\'e}sert}, {Dickinson}, {Diego},
  {Donzelli}, {Dor{\'e}}, {Douspis}, {Dunkley}, {Dupac}, {Efstathiou},
  {En{\ss}lin}, {Eriksen}, {Falgarone}, {Ferri{\`e}re}, {Finelli}, {Forni},
  {Frailis}, {Fraisse}, {Franceschi}, {Galeotta}, {Ganga}, {Ghosh}, {Giard},
  {Giraud-H{\'e}raud}, {Gonz{\'a }lez-Nuevo}, {G{\'o}rski}, {Gregorio},
  {Gruppuso}, {Guillet}, {Hansen}, {Harrison}, {Helou}, {Hern{\'a}ndez-
  Monteagudo}, {Hildebrandt}, {Hivon}, {Hobson}, {Holmes}, {Hornstrup},
  {Huffenberger}, {Jaffe}, {Jaffe}, {Jones}, {Juvela}, {Keih{\"a}nen},
  {Keskitalo}, {Kisner}, {Kneissl}, {Knoche}, {Kunz}, {Kurki-Suonio},
  {Lagache}, {L{\"a}hteenm{\"a}ki}, {Lamarre}, {Lasenby}, {Lawrence}, {Leahy},
  {Leonardi}, {Levrier}, {Liguori}, {Lilje}, {Linden-V{\o}rnle}, {L{\'o}pez-
  Caniego}, {Lubin}, {Mac{\'\i}as-P{\'e}rez}, {Maffei}, {Magalh{\~a}es},
  {Maino}, {Mandolesi}, {Maris}, {Marshall}, {Martin},
  {Mart{\'\i}nez-Gonz{\'a}lez}, {Masi}, {Matarrese}, {Mazzotta}, {Melchiorri},
  {Mendes}, {Mennella}, {Migliaccio}, {Miville-Desch{\^e}nes}, {Moneti},
  {Montier}, {Morgante}, {Mortlock}, {Munshi}, {Murphy}, {Naselsky}, {Nati},
  {Natoli}, {Netterfield}, {Noviello}, {Novikov}, {Novikov}, {Oxborrow},
  {Pagano}, {Pajot}, {Paladini}, {Paoletti}, {Pasian}, {Pearson}, {Perdereau},
  {Perotto}, {Perrotta}, {Piacentini}, {Piat}, {Pietrobon}, {Plaszczynski},
  {Poidevin}, {Pointecouteau}, {Polenta}, {Popa}, {Pratt}, {Prunet}, {Puget},
  {Rachen}, {Reach}, {Rebolo}, {Reinecke}, {Remazeilles}, {Renault},
  {Ricciardi}, {Riller}, {Ristorcelli}, {Rocha}, {Rosset}, {Roudier},
  {Rubi{\~n}o-Mart{\'\i}n}, {Rusholme}, {Sandri}, {Savini}, {Scott}, {Spencer},
  {Stolyarov}, {Stompor}, {Sudiwala}, {Sutton}, {Suur-Uski}, {Sygnet},
  {Tauber}, {Terenzi}, {Toffolatti}, {Tomasi}, {Tristram}, {Tucci}, {Umana},
  {Valenziano}, {Valiviita}, {Van Tent}, {Vielva}, {Villa}, {Wade}, {Wandelt},
  {Zacchei}, \& {Zonca}}]{PlanckXIX15}
{Planck Collaboration}, {Ade}, P.~A.~R., {Aghanim}, N., {et~al.}
  2015{\natexlab{a}}, \aap, 576, A104

\bibitem[{{Planck Collaboration} {et~al.}(2016){Planck Collaboration}, {Ade},
  {Aghanim}, {Alves}, {Arnaud}, {Arzoumanian}, {Ashdown}, {Aumont},
  {Baccigalupi}, {Band ay}, {Barreiro}, {Bartolo}, {Battaner}, {Benabed},
  {Beno{\^\i}t}, {Benoit-L{\'e}vy}, {Bernard}, {Bersanelli}, {Bielewicz},
  {Bock}, {Bonavera}, {Bond}, {Borrill}, {Bouchet}, {Boulanger}, {Bracco},
  {Burigana}, {Calabrese}, {Cardoso}, {Catalano}, {Chiang}, {Christensen},
  {Colombo}, {Combet}, {Couchot}, {Crill}, {Curto}, {Cuttaia}, {Danese},
  {Davies}, {Davis}, {de Bernardis}, {de Rosa}, {de Zotti}, {Delabrouille},
  {Dickinson}, {Diego}, {Dole}, {Donzelli}, {Dor{\'e}}, {Douspis}, {Ducout},
  {Dupac}, {Efstathiou}, {Elsner}, {En{\ss}lin}, {Eriksen},
  {Falceta-Gon{\c{c}}alves}, {Falgarone}, {Ferri{\`e}re}, {Finelli}, {Forni},
  {Frailis}, {Fraisse}, {Franceschi}, {Frejsel}, {Galeotta}, {Galli}, {Ganga},
  {Ghosh}, {Giard}, {Gjerl{\o}w}, {Gonz{\'a}lez-Nuevo}, {G{\'o}rski},
  {Gregorio}, {Gruppuso}, {Gudmundsson}, {Guillet}, {Harrison}, {Helou},
  {Hennebelle}, {Henrot-Versill{\'e}}, {Hern{\'a}ndez-Monteagudo}, {Herranz},
  {Hildebrand t}, {Hivon}, {Holmes}, {Hornstrup}, {Huffenberger}, {Hurier},
  {Jaffe}, {Jaffe}, {Jones}, {Juvela}, {Keih{\"a}nen}, {Keskitalo}, {Kisner},
  {Knoche}, {Kunz}, {Kurki-Suonio}, {Lagache}, {Lamarre}, {Lasenby},
  {Lattanzi}, {Lawrence}, {Leonardi}, {Levrier}, {Liguori}, {Lilje},
  {Linden-V{\o}rnle}, {L{\'o}pez-Caniego}, {Lubin}, {Mac{\'\i}as-P{\'e}rez},
  {Maino}, {Mandolesi}, {Mangilli}, {Maris}, {Martin},
  {Mart{\'\i}nez-Gonz{\'a}lez}, {Masi}, {Matarrese}, {Melchiorri}, {Mendes},
  {Mennella}, {Migliaccio}, {Miville-Desch{\^e}nes}, {Moneti}, {Montier},
  {Morgante}, {Mortlock}, {Munshi}, {Murphy}, {Naselsky}, {Nati},
  {Netterfield}, {Noviello}, {Novikov}, {Novikov}, {Oppermann}, {Oxborrow},
  {Pagano}, {Pajot}, {Paladini}, {Paoletti}, {Pasian}, {Perotto}, {Pettorino},
  {Piacentini}, {Piat}, {Pierpaoli}, {Pietrobon}, {Plaszczynski},
  {Pointecouteau}, {Polenta}, {Ponthieu}, {Pratt}, {Prunet}, {Puget}, {Rachen},
  {Reinecke}, {Remazeilles}, {Renault}, {Renzi}, {Ristorcelli}, {Rocha},
  {Rossetti}, {Roudier}, {Rubi{\~n}o-Mart{\'\i}n}, {Rusholme}, {Sandri},
  {Santos}, {Savelainen}, {Savini}, {Scott}, {Soler}, {Stolyarov}, {Sudiwala},
  {Sutton}, {Suur-Uski}, {Sygnet}, {Tauber}, {Terenzi}, {Toffolatti}, {Tomasi},
  {Tristram}, {Tucci}, {Umana}, {Valenziano}, {Valiviita}, {Van Tent},
  {Vielva}, {Villa}, {Wade}, {Wandelt}, {Wehus}, {Ysard}, {Yvon}, \&
  {Zonca}}]{PlanckXXXV16}
{Planck Collaboration}, {Ade}, P.~A.~R., {Aghanim}, N., {et~al.} 2016, \aap,
  586, A138

\bibitem[{{Planck Collaboration} {et~al.}(2018){Planck Collaboration},
  {Aghanim}, {Akrami}, {Alves}, {Ashdown}, {Aumont}, {Baccigalupi},
  {Ballardini}, {Banday}, {Barreiro}, {Bartolo}, {Basak}, {Benabed}, {Bernard},
  {Bersanelli}, {Bielewicz}, {Bock}, {Bond}, {Borrill}, {Bouchet}, {Boulanger},
  {Bracco}, {Bucher}, {Burigana}, {Calabrese}, {Cardoso}, {Carron}, {Chary},
  {Chiang}, {Colombo}, {Combet}, {Crill}, {Cuttaia}, {de Bernardis}, {de
  Zotti}, {Delabrouille}, {Delouis}, {Di Valentino}, {Dickinson}, {Diego},
  {Dor{\'e}}, {Douspis}, {Ducout}, {Dupac}, {Efstathiou}, {Elsner},
  {En{\ss}lin}, {Eriksen}, {Fantaye}, {Fernandez-Cobos}, {Ferri{\`e}re},
  {Forastieri}, {Frailis}, {Fraisse}, {Franceschi}, {Frolov}, {Galeotta},
  {Galli}, {Ganga}, {G{\'e}nova-Santos}, {Gerbino}, {Ghosh},
  {Gonz{\'a}lez-Nuevo}, {G{\'o}rski}, {Gratton}, {Green}, {Gruppuso},
  {Gudmundsson}, {Guillet}, {Handley}, {Hansen}, {Helou}, {Herranz}, {Hivon},
  {Huang}, {Jaffe}, {Jones}, {Keih{\"a}nen}, {Keskitalo}, {Kiiveri}, {Kim},
  {Krachmalnicoff}, {Kunz}, {Kurki-Suonio}, {Lagache}, {Lamarre}, {Lasenby},
  {Lattanzi}, {Lawrence}, {Le Jeune}, {Levrier}, {Liguori}, {Lilje},
  {Lindholm}, {L{\'o}pez-Caniego}, {Lubin}, {Ma}, {Mac{\'\i}as-P{\'e}rez},
  {Maggio}, {Maino}, {Mandolesi}, {Mangilli}, {Marcos-Caballero}, {Maris},
  {Martin}, {Mart{\'\i}nez-Gonz{\'a}lez}, {Matarrese}, {Mauri}, {McEwen},
  {Melchiorri}, {Mennella}, {Migliaccio}, {Miville-Desch{\^e}nes}, {Molinari},
  {Moneti}, {Montier}, {Morgante}, {Moss}, {Natoli}, {Pagano}, {Paoletti},
  {Patanchon}, {Perrotta}, {Pettorino}, {Piacentini}, {Polastri}, {Polenta},
  {Puget}, {Rachen}, {Reinecke}, {Remazeilles}, {Renzi}, {Ristorcelli},
  {Rocha}, {Rosset}, {Roudier}, {Rubi{\~n}o-Mart{\'\i}n}, {Ruiz-Granados},
  {Salvati}, {Sandri}, {Savelainen}, {Scott}, {Sirignano}, {Sunyaev},
  {Suur-Uski}, {Tauber}, {Tavagnacco}, {Tenti}, {Toffolatti}, {Tomasi},
  {Trombetti}, {Valiviita}, {Van Tent}, {Vielva}, {Villa}, {Vittorio},
  {Wandelt}, {Wehus}, {Zacchei}, \& {Zonca}}]{PlanckXII18}
{Planck Collaboration}, {Aghanim}, N., {Akrami}, Y., {et~al.} 2018, arXiv
  e-prints, arXiv:1807.06212

\bibitem[{{Planck Collaboration} {et~al.}(2015{\natexlab{b}}){Planck
  Collaboration}, {Fermi Collaboration}, {Ade}, {Aghanim}, {Aniano}, {Arnaud},
  {Ashdown}, {Aumont}, {Baccigalupi}, {Banday}, {Barreiro}, {Bartolo},
  {Battaner}, {Benabed}, {Benoit-L{\'e}vy}, {Bernard}, {Bersanelli},
  {Bielewicz}, {Bonaldi}, {Bonavera}, {Bond}, {Borrill}, {Bouchet},
  {Boulanger}, {Burigana}, {Butler}, {Calabrese}, {Cardoso}, {Casandjian},
  {Catalano}, {Chamballu}, {Chiang}, {Christensen}, {Colombo}, {Combet},
  {Couchot}, {Crill}, {Curto}, {Cuttaia}, {Danese}, {Davies}, {Davis}, {de
  Bernardis}, {de Rosa}, {de Zotti}, {Delabrouille}, {D{\'e}sert}, {Dickinson},
  {Diego}, {Digel}, {Dole}, {Donzelli}, {Dor{\'e}}, {Douspis}, {Ducout},
  {Dupac}, {Efstathiou}, {Elsner}, {En{\ss}lin}, {Eriksen}, {Falgarone},
  {Finelli}, {Forni}, {Frailis}, {Fraisse}, {Franceschi}, {Frejsel}, {Fukui},
  {Galeotta}, {Galli}, {Ganga}, {Ghosh}, {Giard}, {Gjerl{\o}w},
  {Gonz{\'a}lez-Nuevo}, {G{\'o}rski}, {Gregorio}, {Grenier}, {Gruppuso},
  {Hansen}, {Hanson}, {Harrison}, {Henrot-Versill{\'e}},
  {Hern{\'a}ndez-Monteagudo}, {Herranz}, {Hildebrandt}, {Hivon}, {Hobson},
  {Holmes}, {Hovest}, {Huffenberger}, {Hurier}, {Jaffe}, {Jaffe}, {Jones},
  {Juvela}, {Keih{\"a}nen}, {Keskitalo}, {Kisner}, {Kneissl}, {Knoche}, {Kunz},
  {Kurki- Suonio}, {Lagache}, {Lamarre}, {Lasenby}, {Lattanzi}, {Lawrence},
  {Leonardi}, {Levrier}, {Liguori}, {Lilje}, {Linden-V{\o}rnle},
  {L{\'o}pez-Caniego}, {Lubin}, {Mac{\'\i}as-P{\'e}rez}, {Maffei}, {Maino},
  {Mandolesi}, {Maris}, {Marshall}, {Martin}, {Mart{\'\i}nez- Gonz{\'a}lez},
  {Masi}, {Matarrese}, {Mazzotta}, {Melchiorri}, {Mendes}, {Mennella},
  {Migliaccio}, {Miville-Desch{\^e}nes}, {Moneti}, {Montier}, {Morgante},
  {Mortlock}, {Munshi}, {Murphy}, {Naselsky}, {Natoli}, {N{\o}rgaard-Nielsen},
  {Novikov}, {Novikov}, {Oxborrow}, {Pagano}, {Pajot}, {Paladini}, {Paoletti},
  {Pasian}, {Perdereau}, {Perotto}, {Perrotta}, {Pettorino}, {Piacentini},
  {Piat}, {Plaszczynski}, {Pointecouteau}, {Polenta}, {Popa}, {Pratt},
  {Prunet}, {Puget}, {Rachen}, {Reach}, {Rebolo}, {Reinecke}, {Remazeilles},
  {Renault}, {Ristorcelli}, {Rocha}, {Roudier}, {Rusholme}, {Sandri}, {Santos},
  {Scott}, {Spencer}, {Stolyarov}, {Strong}, {Sudiwala}, {Sunyaev}, {Sutton},
  {Suur-Uski}, {Sygnet}, {Tauber}, {Terenzi}, {Tibaldo}, {Toffolatti},
  {Tomasi}, {Tristram}, {Tucci}, {Umana}, {Valenziano}, {Valiviita}, {Van
  Tent}, {Vielva}, {Villa}, {Wade}, {Wandelt}, {Wehus}, {Yvon}, {Zacchei}, \&
  {Zonca}}]{PlanckXXVIII15}
{Planck Collaboration}, {Fermi Collaboration}, {Ade}, P.~A.~R., {et~al.}
  2015{\natexlab{b}}, \aap, 582, A31

\bibitem[{{Pon} {et~al.}(2014){Pon}, {Johnstone}, {Bally}, \& {Heiles}}]{Pon14}
{Pon}, A., {Johnstone}, D., {Bally}, J., \& {Heiles}, C. 2014, \mnras, 441,
  1095

\bibitem[{{Pon} {et~al.}(2016){Pon}, {Ochsendorf}, {Alves}, {Bally}, {Basu}, \&
  {Tielens}}]{Pon16}
{Pon}, A., {Ochsendorf}, B.~B., {Alves}, J., {et~al.} 2016, \apj, 827, 42

\bibitem[{{Puspitarini} {et~al.}(2014){Puspitarini}, {Lallement}, {Vergely}, \&
  {Snowden}}]{Puspitarini14}
{Puspitarini}, L., {Lallement}, R., {Vergely}, J.~L., \& {Snowden}, S.~L. 2014,
  \aap, 566, A13

\bibitem[{{Remy} {et~al.}(2017){Remy}, {Grenier}, {Marshall}, \&
  {Casandjian}}]{Remy17}
{Remy}, Q., {Grenier}, I.~A., {Marshall}, D.~J., \& {Casandjian}, J.~M. 2017,
  \aap, 601, A78

\bibitem[{{Reynolds} \& {Ogden}(1979)}]{Reynolds1979}
{Reynolds}, R.~J. \& {Ogden}, P.~M. 1979, \apj, 229, 942

\bibitem[{{Richings} {et~al.}(2014){Richings}, {Schaye}, \&
  {Oppenheimer}}]{Richings14}
{Richings}, A.~J., {Schaye}, J., \& {Oppenheimer}, B.~D. 2014, \mnras, 440,
  3349

\bibitem[{{Russeil} {et~al.}(2003){Russeil}, {Juvela}, {Lehtinen}, {Mattila},
  \& {Paatero}}]{Russeil2003}
{Russeil}, D., {Juvela}, M., {Lehtinen}, K., {Mattila}, K., \& {Paatero}, P.
  2003, \aap, 409, 135

\bibitem[{{Schlegel} {et~al.}(1998){Schlegel}, {Finkbeiner}, \&
  {Davis}}]{SFD1998}
{Schlegel}, D.~J., {Finkbeiner}, D.~P., \& {Davis}, M. 1998, \apj, 500, 525

\bibitem[{{Snowden} {et~al.}(1995){Snowden}, {Burrows}, {Sanders},
  {Aschenbach}, \& {Pfeffermann}}]{Snowden95}
{Snowden}, S.~L., {Burrows}, D.~N., {Sanders}, W.~T., {Aschenbach}, B., \&
  {Pfeffermann}, E. 1995, \apj, 439, 399

\bibitem[{{Snowden} {et~al.}(1994){Snowden}, {McCammon}, {Burrows}, \&
  {Mendenhall}}]{Snowden94}
{Snowden}, S.~L., {McCammon}, D., {Burrows}, D.~N., \& {Mendenhall}, J.~A.
  1994, \apj, 424, 714

\bibitem[{{Sofue} {et~al.}(2019){Sofue}, {Nakanishi}, \& {Ichiki}}]{Sofue19}
{Sofue}, Y., {Nakanishi}, H., \& {Ichiki}, K. 2019, \mnras, 485, 924

\bibitem[{{Soler} {et~al.}(2018){Soler}, {Bracco}, \& {Pon}}]{Soler18}
{Soler}, J.~D., {Bracco}, A., \& {Pon}, A. 2018, \aap, 609, L3

\bibitem[{{Voss} {et~al.}(2010){Voss}, {Diehl}, {Vink}, \& {Hartmann}}]{Voss10}
{Voss}, R., {Diehl}, R., {Vink}, J.~S., \& {Hartmann}, D.~H. 2010, \aap, 520,
  A51

\bibitem[{{Wakker} {et~al.}(2008){Wakker}, {York}, {Wilhelm}, {Barentine},
  {Richter}, {Beers}, {Ivezi{\'c}}, \& {Howk}}]{Wakker2008}
{Wakker}, B.~P., {York}, D.~G., {Wilhelm}, R., {et~al.} 2008, \apj, 672, 298

\bibitem[{{Welsh} {et~al.}(2005){Welsh}, {Sallmen}, \& {Jelinsky}}]{Welsh05}
{Welsh}, B.~Y., {Sallmen}, S., \& {Jelinsky}, S. 2005, \aap, 440, 547

\bibitem[{{Xu} \& {Han}(2019)}]{Xu19}
{Xu}, J. \& {Han}, J.~L. 2019, \mnras, 486, 4275

\bibitem[{{Zucker} {et~al.}(2019){Zucker}, {Speagle}, {Schlafly}, {Green},
  {Finkbeiner}, {Goodman}, \& {Alves}}]{Zucker2019}
{Zucker}, C., {Speagle}, J.~S., {Schlafly}, E.~F., {et~al.} 2019, \apj, 879,
  125

\end{thebibliography}

\appendix

\section{X-ray optical depth}
\label{ap:tauX}

\begin{figure*}[ht]
\centering
\includegraphics[width=17cm]{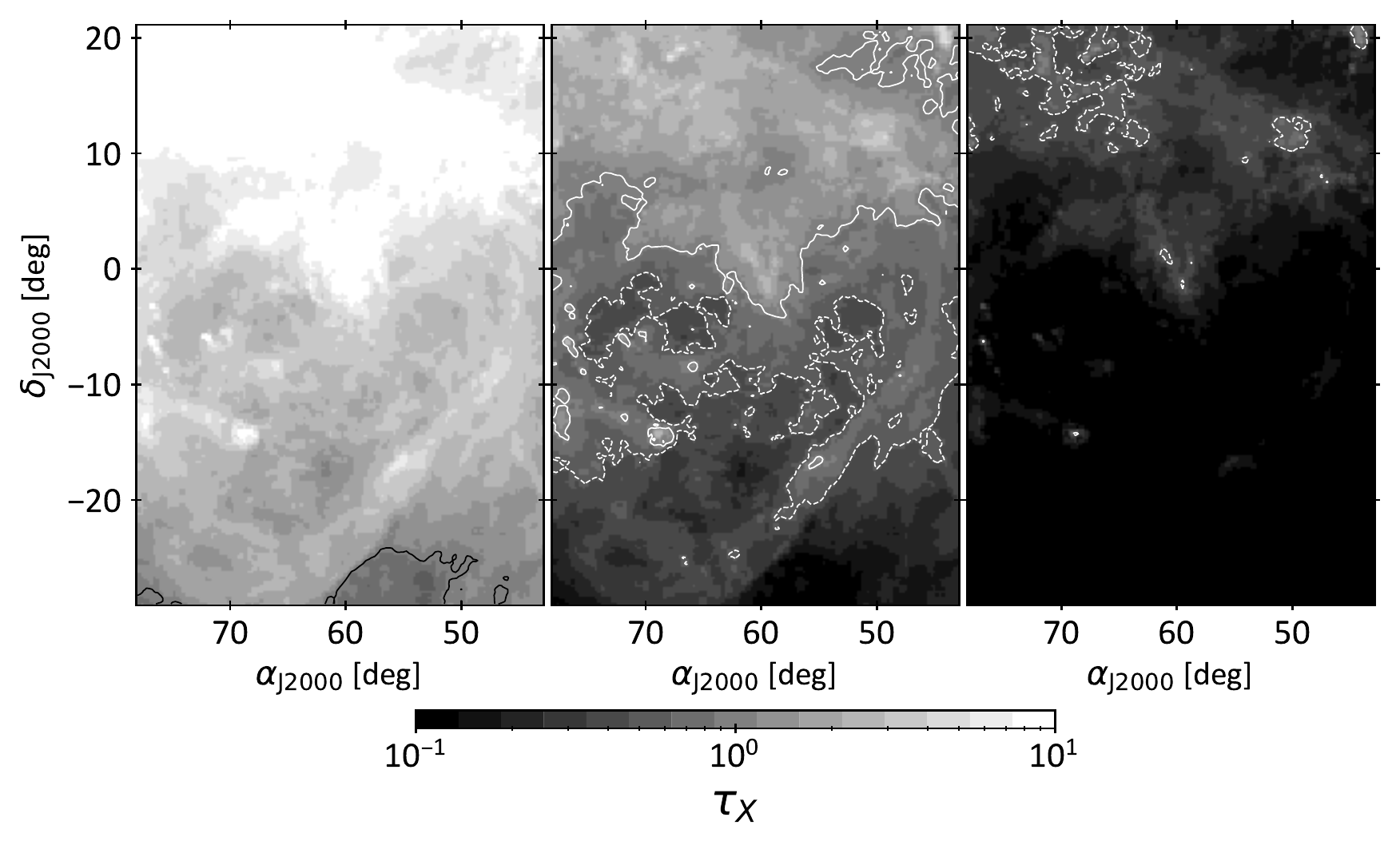}
\caption{X-ray optical depth for the total gas in the 0.25 keV (\textit{left}), 0.75 keV (\textit{middle}), and 1.5 keV (\textit{right}) energy bands. The contours outline the $\tau_{\rm X}=0.5$ (dashed) and $\tau_{\rm X}=1$ (solid) levels.}
\label{fig:tauX}
\end{figure*}

In order to analyse the X-ray emission from the superbubble, we have computed the X-ray optical depth maps using
\begin{equation}
\tau_X(E_{\rm X}) = \sigma(E_{\rm X}, N_{\rm H}) \times N_{\rm H}
,\end{equation}
where $E_{\rm X}$ is the X-ray band, $\sigma$ the energy-band-averaged photoelectric absorption cross section from \citet{Snowden94} and \nh the gas column densities inferred in the different gas phases from our combined \hi, CO, dust, and \g-ray study. The optical depth maps obtained for the total gas column densities $\tau_{tot\,X}$ and for the three ROSAT energy bands are displayed in Figure \ref{fig:tauX}.

\begin{figure}[ht]
\resizebox{\hsize}{!}{\includegraphics{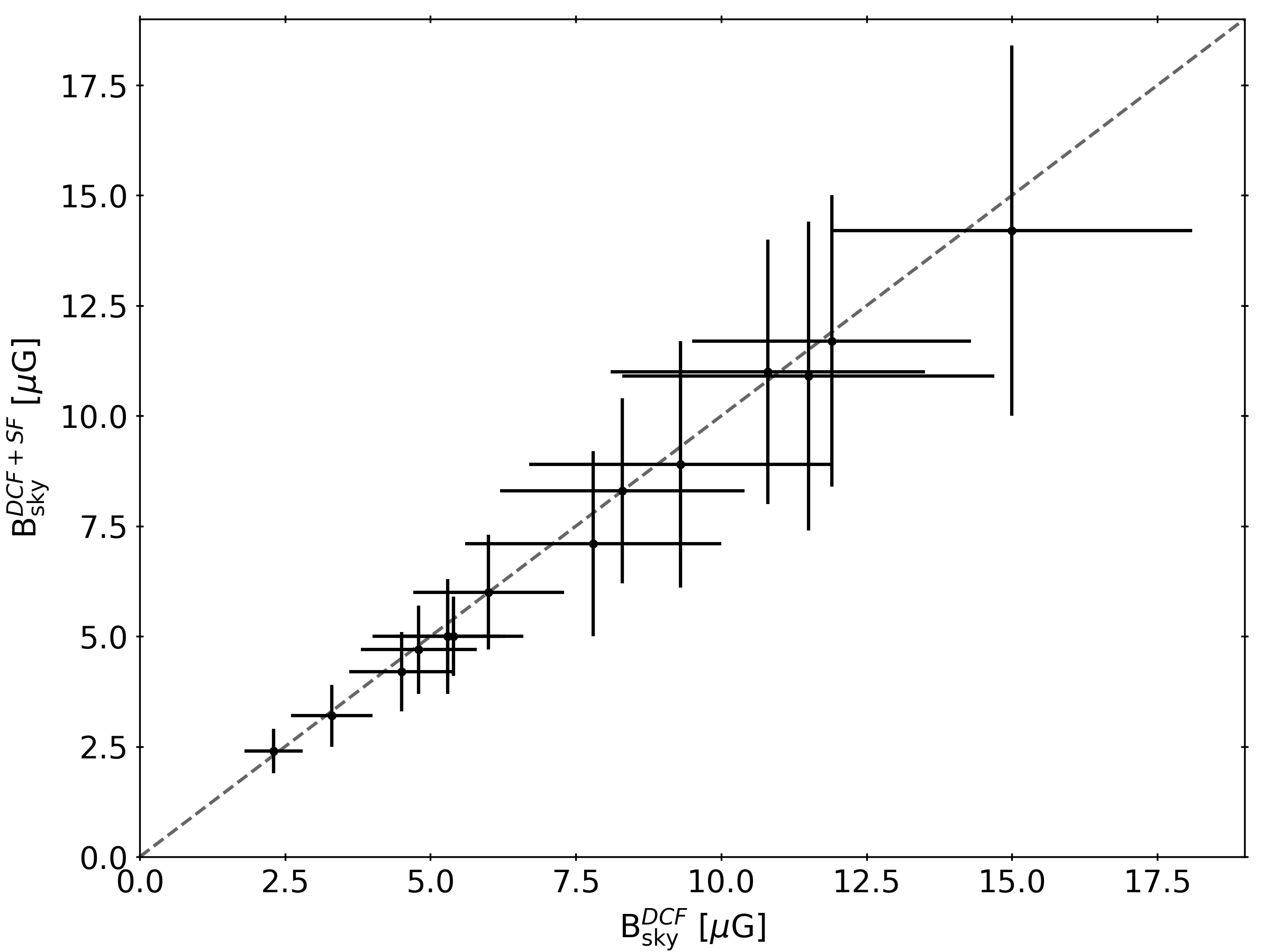}}
\caption{Comparison of the plane-of-sky magnetic field strengths with two different methods to estimate the dispersion of the polarisation angles: directly from the Stokes parameters, \bsky, and using the structure function method, $B_{\rm sky}^{DCF+SF}$.}
\label{fig:Bcomp}
\end{figure}

\section{Angular dispersion of the magnetic field}
\label{ap:polar}

To derive the angular dispersion of the magnetic field, we have first used the direct method described in Eqs. (D.5) and (D.11) of \citet{PlanckXXXV16}, 
\begin{equation*}
\varsigma_{\psi} = \sqrt{\langle(\Delta \psi_x)^{2}\rangle},
\end{equation*}
and
\begin{equation*}
\Delta \psi_x = \psi(\bm{x})- \langle\psi\rangle = \frac{1}{2} \mathrm{arctan}(Q\langle U \rangle - U\langle Q \rangle, Q\langle Q \rangle + U\langle U \rangle),
\end{equation*}

where $\psi(\bm{x})$ denotes the polarisation angle at the position $\bm{x}$ in the sky, $\langle ... \rangle$ is the average over the circle with a diameter of 2\degr\ , arctan(sin, cos) is the arc-tangent function that solves the $\pi$ ambiguity taking into account the sign of the cosine, and $Q$ and $U$ are the Stokes parameters.

The second method relies on the structure function of polarisation angles, $\mathcal{S}_2(\delta)$, defined in \citet{PlanckXIX15} as 

\begin{equation*}
\mathcal{S}_2(\delta) = \left\langle \left( \frac{1}{N} \sum_{i=1}^{N}(\Delta \psi_{xi})^{2}\right)^{1/2} \right\rangle,
\end{equation*}
and
\begin{equation*}
\Delta \psi_{xi} = \psi(\bm{x})- \psi(\bm{x}+\bm{\delta}_i) = \frac{1}{2} \mathrm{arctan}(Q_i U_x - U_i Q_x, Q_i Q_x + U_i U_x),
\end{equation*}

where $\Delta \psi_{xi}$ is the angle difference between the polarisation at the position in the sky $\bm{x}$ (central pixel) and the polarisation at a position displaced by the vector $\bm{\delta}_i$. The sum is over an annulus around the central pixel $\bm{x}$ of radius $\delta=\lvert\bm{\delta}\rvert$ (the lag), width $\Delta \delta$ and containing $N$ pixels. We chose a width $\Delta \delta = \delta$ as advised in \citet{PlanckXIX15}. The result was then averaged over all the positions $\bm{x}$ in the circle of 2\degr\ diameter, leaving only a dependence on $\delta$ .

\citet{Hildebrand09} considered a magnetic field composed of a field with a large-scale structure, $\bm{B}_0$, and a turbulent component, $\bm{B}_t$. Assuming that both contributions are statistically independent, they made the approximation  $\mathcal{S}_2^2(\delta) = s^2+ m\delta^2$, the first term being the contribution of $\bm{B}_t$ and the second term the linear contribution of the smoothly varying $\bm{B}_0$. We derived the intercept $s$ by fitting the structure function with this model on the range $\delta=[1\degr,2\degr]$, above the 40' resolution of the data. $s$ was then linked to the plane-of-sky magnetic field, taking into account the \citet{Cho16} correction, through 

\begin{equation*}
\rm{B^{DCF+SF}_{\rm sky} = \sqrt{4 \pi \rho} \frac{\delta V_{c} \sqrt{2-s^2}}{s}}
\end{equation*}

We consider here small regions compared to the \Planck beam size, however, which implies few independent vectors in our apertures with 2$\degr$ diameter. As a consequence, the structure function method that probes the magnetic field at different angular scales gives very similar results to the direct method that only probes the average, as shown in  Figure \ref{fig:Bcomp}. Only the magnetic field strengths from the direct method are displayed in Table \ref{tab:BDCF}.

\end{document}